\documentclass[preprint,12pt]{elsarticle}

\usepackage{amssymb}
\usepackage{amsmath}
\usepackage{enumitem}
\usepackage{hyperref}
\usepackage{float}
\usepackage{xurl}
\usepackage{listings}
\usepackage{booktabs}
\usepackage{multirow}
\usepackage[T1]{fontenc}
\journal{Information and Software Technology}

\begin{document}

\begin{frontmatter}



\title{The Myth of Immutability: A Multivocal Review on Smart Contract Upgradeability} 


\author[label1]{Ilham Qasse} 
\ead{ilham20@ru.is}
\author[label2]{Isra M. Ali\fnmark[1]} 
\ead{https://orcid.org/0000-0002-2689-348X}
\author[label3]{Nafisa Ahmed\fnmark[1]} 
\ead{nafisa.abdelmutalab-ali-ahmed@polymtl.ca}
\author[label1,label3]{Mohammad Hamdaqa}
\ead{mhamdaqa@polymtl.ca}
\ead{mhamdaqa@ru.is}
\author[label1]{Björn Þór Jónsson} 
\ead{bjorn@ru.is}

\fntext[1]{These authors contributed equally to this work.}
\affiliation[label1]{organization={Department of Computer Science, Reykjavik University},
            city={Reykjavik},
            country={Iceland}}
\affiliation[label2]{organization={College of Science and Engineering, Hamad Bin Khalifa University},
            city={Doha},
            country={Qatar}}
\affiliation[label3]{organization={Department of Computer Engineering and Software Engineering, Polytechnique Montreal},
            city={Montreal},
            country={Canada}}
\begin{abstract}
\textbf{Background:}
The immutability of smart contracts on blockchain platforms like Ethereum promotes security and trustworthiness but presents challenges for updates, bug fixes, or adding new features post-deployment. These limitations can lead to vulnerabilities and outdated functionality, impeding the evolution and maintenance of decentralized applications. Despite various upgrade mechanisms proposed in academic research and industry, a comprehensive analysis of their trade-offs and practical implications is lacking.

\noindent\textbf{Aims:}
This study aims to systematically identify, classify, and evaluate existing smart contract upgrade mechanisms, bridging the gap between theoretical concepts and practical implementations. It introduces standardized terminology and evaluates the trade-offs of different approaches using software quality attributes.

\noindent\textbf{Methods:}
We conducted a Multivocal Literature Review (MLR) to analyze upgrade mechanisms from both academic research and industry practice. We first establish a unified definition of smart contract upgradeability and identify core components essential for understanding the upgrade process. Based on this definition, we classify existing methods into full upgrade and partial upgrade approaches, introducing standardized terminology to harmonize the diverse terms used in the literature. We then characterize each approach and assess its benefits and limitations using software quality attributes such as complexity, flexibility, security, and usability.

\noindent\textbf{Results:}
The analysis highlights significant trade-offs among upgrade mechanisms, providing valuable insights into the benefits and limitations of each approach. These findings guide developers and researchers in selecting mechanisms tailored to specific project requirements.

\noindent\textbf{Conclusions:}
By offering a comprehensive evaluation that bridges theoretical concepts and practical implementations, this study contributes to the development of more resilient, adaptable, and secure decentralized systems.
\end{abstract}


\begin{highlights}
\item Defines smart contract upgradeability, uniting varied perspectives in the field.
\item Presents a unified taxonomy and classification of smart contract upgrade methods.
\item Examines the impact of upgrade approaches on core smart contract components.
\item Analyzes the benefits and limitations of upgrade methods based on software quality attributes.
\item Provides practical insights, best practices, and future research directions.
\end{highlights}

\begin{keyword}
Smart Contract \sep Upgradeability \sep Blockchain Technology \sep Immutability \sep Multivocal Literature Review \sep Ethereum \sep Software Maintenance \sep Proxy Patterns \sep Decentralized Applications (DApps) \sep Upgrade Patterns \sep Software Quality Attributes



\end{keyword}

\end{frontmatter}



\section{Introduction}
\label{sec1}

Software systems constantly evolve to fix bugs, add new features, and adapt to changing user needs and technological environments~\cite{Lehman1980, Sampaio2023}. This ongoing evolution is fundamental to software engineering, ensuring that applications stay relevant, secure, and efficient over time~\cite{Zheng2020}. Traditional software upgrade methods allow developers to update systems smoothly, minimizing user disruptions and maintaining system stability~\cite{Buterin2014}. However, the advent of blockchain technology and smart contracts has introduced a new paradigm in software evolution, fundamentally challenging traditional notions of upgradeability.

Smart contracts are self-executing programs stored on a blockchain, designed to uphold the principles of immutability to ensure transparency, security, and trustworthiness~\cite{Nikolic2018}. Once deployed, a smart contract's code is permanent and cannot be altered, posing a significant challenge: \textit{How can developers upgrade smart contracts without undermining the immutability that ensures their reliability}~\cite{Antonino2022}? This inherent tension between the need for software evolution and the immutable nature of smart contracts has become a critical issue in blockchain development.

High-profile incidents, such as the infamous DAO attack~\cite{Siegel2016, Grishchenko2018},  where a vulnerability in a smart contract was exploited to drain over \$60 million worth of Ether, highlight the urgency of this problem. Vulnerabilities in deployed contracts expose the limitations of immutability, where the inability to make post-deployment modifications prevents fixing errors, adding new features, and meeting new regulations~\cite{Atzei2017, Chen2020}. These challenges are especially significant in Ethereum, the most widely used platform for smart contracts, since it supports a large ecosystem of decentralized applications (DApps) and decentralized finance (DeFi) protocols.

Several studies have explored aspects of smart contract upgradeability~\cite{ebrahimi2024large, salehi2022not, bodell2023proxy, qasse2024immutable, bui2021evaluating}. Salehi et al.~\cite{salehi2022not} summarized and evaluated six upgradeability patterns, developing a framework to measure the prevalence of these patterns on the Ethereum blockchain. Bui et al.~\cite{bui2021evaluating} analyzed the need for upgradability in smart contracts, thoroughly reviewed existing upgradeable patterns, identified their main differences and limitations, and introduced a new pattern to address them. However, these studies are often limited in scope, focusing on specific aspects such as the evaluation of certain patterns or the introduction of new solutions. There is still a lack of comprehensive analysis that integrates academic research and industry practices, resulting in inconsistent definitions and classifications and restricting the development of standardized, broadly applicable approaches.

To bridge this gap, we conduct a \textit{Multivocal Literature Review} (MLR) that consolidates findings from both peer-reviewed academic sources and grey literature, such as technical blogs, reports, and community forums. The MLR approach is uniquely suited for integrating theoretical and practical perspectives, providing a holistic view that can inform standardized upgrade practices~\cite{Cano2021}. 
The main Research Objectives (ROs) of this study are to:

\begin{enumerate}
    \item \textbf{RO1:} Classify existing approaches for upgrading Ethereum smart contracts, presenting a clear taxonomy that includes diverse methods from theory and practice. 
\item \textbf{RO2:} Analyze the characteristics of each upgrading approach, focusing on essential smart contract components. 
\item \textbf{RO3:} Assess the benefits and limitations of each approach in relation to standard software quality attributes. 
\end{enumerate}

To achieve these objectives, we systematically searched major academic databases and grey literature platforms to capture a broad spectrum of perspectives. Employing thematic analysis and framework analysis, we extracted and synthesized data to comprehensively understand smart contract upgradeability~\cite{Braun2006, Ritchie1994}.

This paper contributes to software engineering and blockchain technology by (i) introducing a comprehensive definition of smart contract upgradeability, addressing differing perspectives in the field; (ii) establishing a standardized taxonomy of upgrade mechanisms, unifying terminology to enhance clarity and communication; (iii) analyzing and classifying the characteristics of upgrade mechanisms, providing insights into their impact on core smart contract components; (iv) evaluating the benefits and limitations of each approach based on software quality attributes, offering practical guidance for developers; and (v) highlighting best practices and common pitfalls, and outlining future research directions, emphasizing the need for empirical studies on security, trust, and lifecycle management of upgrade mechanisms.

The remainder of this paper is organized as follows: Section~\ref{BG} provides background information on smart contracts and their key characteristics, with a focus on Ethereum. Section~\ref{sec:methodology} explains the research methodology, including the search strategy, source selection, and data extraction. In Section~\ref{sec:systematization}, we systematize the main ideas related to upgradeability and present our unified definition. Section~\ref{sec:RQ1} covers RQ1 by classifying current upgrade methods. Section~\ref{sec:RQ2} addresses RQ2 by analyzing the features of each method. Section~\ref{sec:RQ3} covers RQ3 by evaluating the pros and cons of each method based on software quality factors. Section~\ref{sec:discussion} discusses the findings, including governance models and lifecycle management. Section~\ref{sec:threats} explains threats to validity. Finally, Section~\ref{sec:conclusion} concludes the paper and suggests future research directions.
\section{Background}
\label{BG}

Smart contracts are self-executing contracts, with the terms of the agreement directly written into code which operates on blockchain technology. They automatically enforce and execute contractual obligations when predefined conditions are met, eliminating the need for intermediaries~\cite{Blanco2023}. This automation enhances efficiency and reduces the potential for disputes, making smart contracts a significant innovation in decentralized systems. Their importance lies in their ability to provide trusted transactions in trustless\footnote{The term 'trustless' refers to systems where participants do not need to trust each other or a central authority, as the system's rules and outcomes are enforced automatically by technology, such as blockchain.} settings, streamline processes, and reduce administrative costs across various sectors, including finance, supply chain management, and healthcare~\cite{So2020, Zheng2020}.

\subsection{Core Characteristics and Applications}

Smart contracts possess several key characteristics that are essential for their functionality:

\begin{itemize}
    \item \textbf{Transparency}: All smart contract transactions are recorded on a blockchain, allowing all parties to verify the contract's execution and state. This transparency fosters trust among users~\cite{Gursoy2020, Cano2021}.
    \item \textbf{Immutability}: Smart contracts cannot be altered once deployed, ensuring that the agreed-upon terms remain intact. This characteristic prevents tampering and enhances security, although it also poses challenges for updates~\cite{Vacca2023}.
    \item \textbf{Autonomy}: Smart contracts operate independently without human intervention once they are deployed. This autonomy reduces the risk of human error and increases efficiency~\cite{Reshi2023}.
    \item \textbf{Security}: Smart contracts leverage cryptographic techniques to secure transactions and data. The decentralized nature of blockchain technology further enhances security by eliminating single points of failure~\cite{Godoy2022}.
\end{itemize}

These characteristics contribute to the reliability and effectiveness of smart contracts in various applications. For example, in Decentralized Finance (DeFi), smart contracts facilitate automated trading, lending, and borrowing without intermediaries, enhancing efficiency and reducing costs~\cite{Zheng2020, Khan2021}. In Supply Chain Management, they enable transparent tracking of goods and services, minimizing disputes and improving accountability~\cite{Reshi2023, Gursoy2020}. Additionally, voting systems utilize smart contracts to create secure and transparent processes, ensuring accurate vote counts and preventing tampering~\cite{Gursoy2020,soud2020trustvote}.

\subsection{Technical Considerations and Gas Costs}

Gas is a unit of measurement for the computational effort required to execute operations within a blockchain network, such as Ethereum. Each operation performed by a smart contract consumes a specific amount of gas, which users pay for in the network's native cryptocurrency (e.g., Ether on Ethereum). Gas costs play a critical role in influencing the efficiency and usability of smart contracts, as high fees may deter users from engaging with blockchain applications~\cite{Chen2017}. Developers adopt strategies such as efficient code design and optimized data structures to minimize gas costs~\cite{Chen2017, Khan2021}.

Another technical consideration is the Self-Destruct Mechanism (\texttt{SELF\break DESTRUCT} operation). This operation allows a smart contract to remove its code and state from the blockchain, freeing up storage space. While useful for managing outdated contracts, it risks permanently losing state data, which can impact users relying on that information~\cite{Vacca2023}.

Execution Flow and Limitations involve managing function calls that consume gas and must stay within network-imposed gas limits. Exceeding these limits results in "Out of Gas" errors, causing transaction failures~\cite{Zheng2020, Vacca2023}. Proper planning and code optimization are essential to prevent such issues.

\subsection{Security Considerations}

Security is paramount for smart contract functionality due to potential vulnerabilities, which could result in significant financial losses. Error Handling and Fallback Functions are crucial for maintaining contract stability. Fallback functions are triggered when a contract receives Ether without data or when an invalid function is called, helping manage unexpected conditions~\cite{Zheng2020}. Mechanisms like \texttt{require}, \texttt{assert}, and \texttt{revert} ensure that contracts behave as intended and can recover gracefully from errors~\cite{Vacca2023}.

Reentrancy Attacks pose a significant security risk, as seen in the DAO incident, where an attacker exploited reentrancy to repeatedly call back into the contract before its execution was complete, resulting in substantial financial losses~\cite{He2023}. Implementing best practices such as the "checks-effects-interactions" pattern is essential for minimizing such risks~\cite{He2023}. This pattern improves security by structuring contract functions first to validate inputs ('checks'), then update the contract state ('effects'), and only afterwards perform external calls ('interactions'). By ensuring state changes occur before any external interactions, this approach prevents malicious contracts from exploiting reentrant calls to manipulate the contract's state.

\subsection{Adaptability and Immutability}
The balance between adaptability and immutability is a major challenge for smart contract developers. Immutability ensures trust and security by preventing changes after deployment, but it also restricts the ability to update contracts or fix errors. This limitation can impede the long-term adaptability of smart contracts in dynamic environments where changes are often necessary~\cite{Vacca2023, Antonino2022}. Innovative upgrade mechanisms are crucial for maintaining functionality while preserving the core principles of blockchain, such as Transparency and user trust.

\subsection{Focus on Ethereum}

This study focuses on Ethereum due to its widespread adoption as the leading platform for smart contract development. Ethereum has a large, active community that provides extensive support for developers and researchers, contributing valuable resources and discussions~\cite{oliva2020exploratory, connors2022review, hamdaqa2022icontractml}. This strong community presence and the platform's open-source nature make Ethereum an ideal choice for examining upgrade mechanisms through a comprehensive Multivocal Literature Review (MLR). The mature ecosystem of Ethereum and the availability of diverse academic and grey literature further support its selection for this study.

\section{Methodology}
\label{sec:methodology}

\begin{figure}[t!]
    \centering
\includegraphics[scale=0.7]{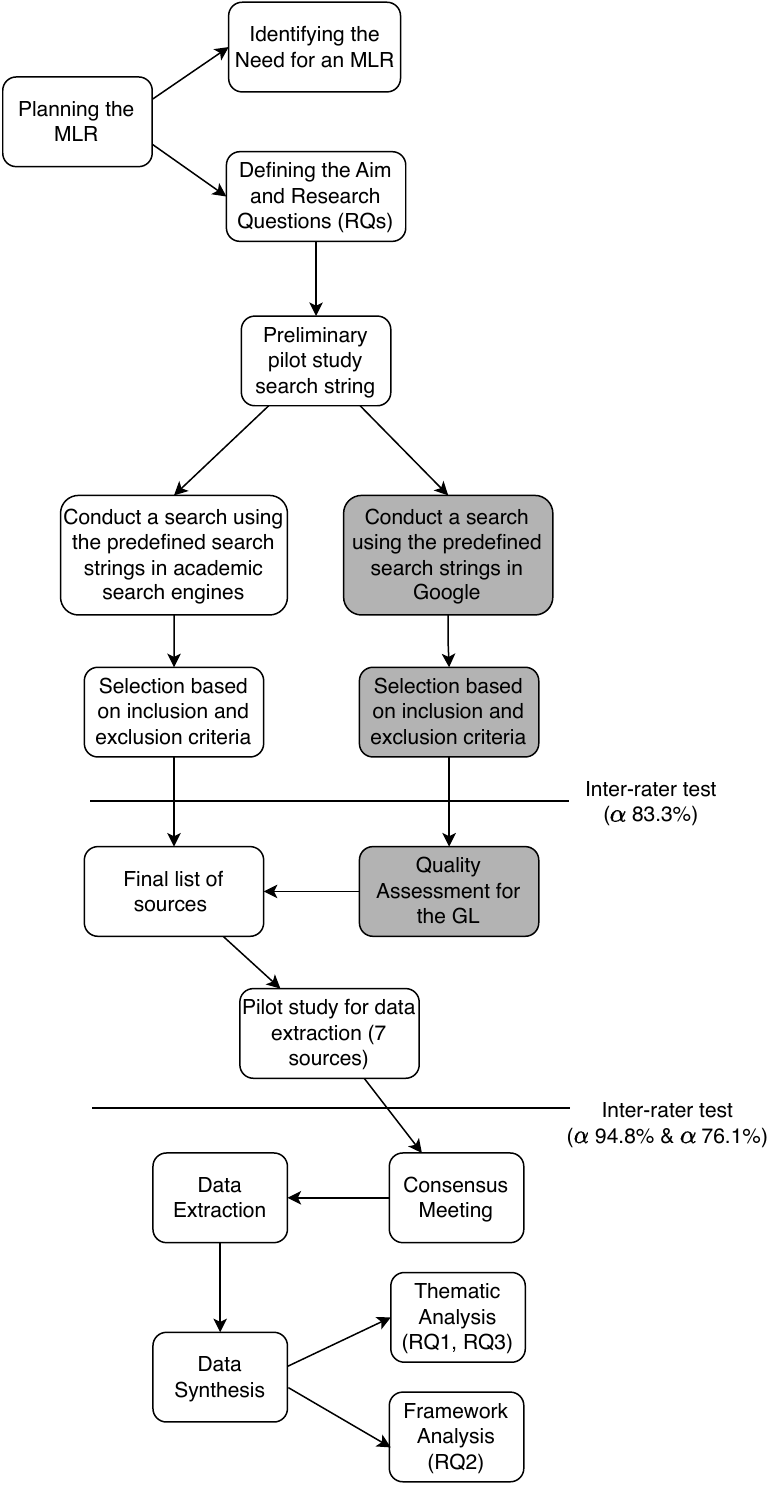} 
    \caption{Outline for research methodology}
    \label{fig:Meth}
\end{figure}

This section presents the methodology for this MLR, following the guidelines outlined by Garousi et al.~\cite{Garousi2019} for conducting Multivocal Literature Reviews (MLRs) and integrating grey literature in software engineering. This structured approach ensures a comprehensive analysis of smart contract upgrade mechanisms by combining academic literature and grey literature (GL). Figure~\ref{fig:Meth} summarizes the methodology, highlighting key steps such as source selection, quality assessment, and data extraction, which are described in detail in the remainder of this section.
\subsection{Planning the Multivocal Literature Review (MLR)}
The first step in the MLR process, as outlined by Garousi et al.~\cite{Garousi2019}, is the planning phase. This involves defining the need for an MLR and formulating clear research questions to guide the review. This section details the rationale for this approach and the research questions that shape the study.

\subsubsection{Identifying the Need for an MLR}
The MLR methodology was chosen over a traditional Systematic Literature Review (SLR) methodology due to the limited availability of peer-reviewed studies on smart contract upgradeability. MLRs integrate both peer-reviewed and grey literature (GL), providing a comprehensive understanding of emerging fields where industry practices often outpace academic research~\cite{Garousi2019}. Grey literature, including technical blogs, white papers, tutorials, and community forum discussions, offers current and practical insights that may not yet be reflected in academic publications. By incorporating GL, we aim to capture both theoretical perspectives and real-world practices in smart contract upgrade mechanisms, ensuring a holistic view of the subject.

\subsubsection{Defining the Aim and Research Questions (RQs)}
The primary aim of this MLR is to explore, classify, and assess the methods for upgrading smart contracts, integrating perspectives from both academia and industry. The following research questions were formulated to address core aspects of smart contract upgradeability:

\begin{itemize}
    \item \textbf{RQ1:} What are the existing approaches for upgrading Ethereum smart contracts? This question identifies and classifies the diverse methods and techniques used to upgrade smart contracts.
    \item \textbf{RQ2:} What are the characteristics of each upgrading approach? This question investigates the key characteristics of each approach based on core smart contract components, such as address preservation, logic upgrade scope, and storage management.
    \item \textbf{RQ3:} What are the benefits and limitations of each approach? This question evaluates the strengths and weaknesses of each approach based on software quality attributes such as flexibility, security, and scalability.
\end{itemize}

\subsection{Search Strategy}
The Search Strategy outlines the process of identifying key terms, conducting pilot testing, and developing the final search string to guide the literature search for this MLR.
\subsubsection{Keyword Selection}
We broke down the research questions into specific terms to develop a comprehensive set of keywords. Keywords were categorized into Ethereum-specific terms, upgrade-related terms, technical methods and patterns, and criteria to capture discussions on benefits and limitations. This structured approach to keyword selection ensured that both technical and practical aspects of smart contract upgradeability were thoroughly explored.

\begin{itemize}
    \item \textbf{Ethereum-Specific Terms:} ``Ethereum'', ``smart contract'', ``Solidity'', ``contract*''
    \item \textbf{Upgrade-Related Terms:} ``upgrade'', ``update*'', ``mutability*'', ``immutability'', ``migration*''
    \item \textbf{Methods and Patterns:} ``pattern'', ``architecture'', ``approach'', ``method'', ``proxy'', ``framework'', ``strategy''
    \item \textbf{Benefits and Limitations:} ``security'', ``challenge*'', ``smell*'', ``issue*'', ``best practice'', ``benefit'', ``advantage*'', ``limitation*'', ``disadvantage*'', ``pro*'', ``con*''
\end{itemize}

The `*` wildcard is used to capture variations of a root term, ensuring the inclusion of related terms in the search results. For instance, ``contract*'' retrieves ``contract'', ``contracts'' and ''contractual'', while ``challenge*'' captures ``challenge'' and ``challenges''. Terms without the wildcard are already comprehensive and do not have meaningful variations relevant to the study. For example, ``Ethereum'' does not require a wildcard as there are no common variants.

\subsubsection{Final Search String}
We conducted pilot testing of the search string on a subset of databases and grey literature sources to assess its effectiveness. The pilot involved running the search string on IEEE Xplore and Google, reviewing the first 50 results to evaluate relevance. During pilot testing, we found that terms like ``security'' and ``smell'' yielded results related to general smart contract vulnerabilities rather than upgrade mechanisms. Therefore, these terms were removed to focus the search on upgradeability topics.

The following search string was developed to integrate the identified keywords, as well as the outcome of the pilot test:
\begin{lstlisting}[basicstyle=\footnotesize]
("Ethereum" OR "smart contract*" OR "solidity" OR "contract*") AND 
("upgrade*" OR "update*" OR "mutability*" OR "migration*") AND 
("pattern" OR "approach" OR "method" OR "proxy" OR "strategy") AND 
("issue" OR "challenge" OR "best practice" OR "benefit*" OR "advantage*" OR 
"limitation*" OR "disadvantage*" OR "pro*" OR "con*")
\end{lstlisting}

The search string was adjusted as necessary to suit the syntax and indexing terms of each academic database. For grey literature searches on Google, parentheses were minimized to accommodate Google's search algorithms.

\subsection{Source Selection}
\begin{figure}[t]
    \centering
    \includegraphics[width=\textwidth]{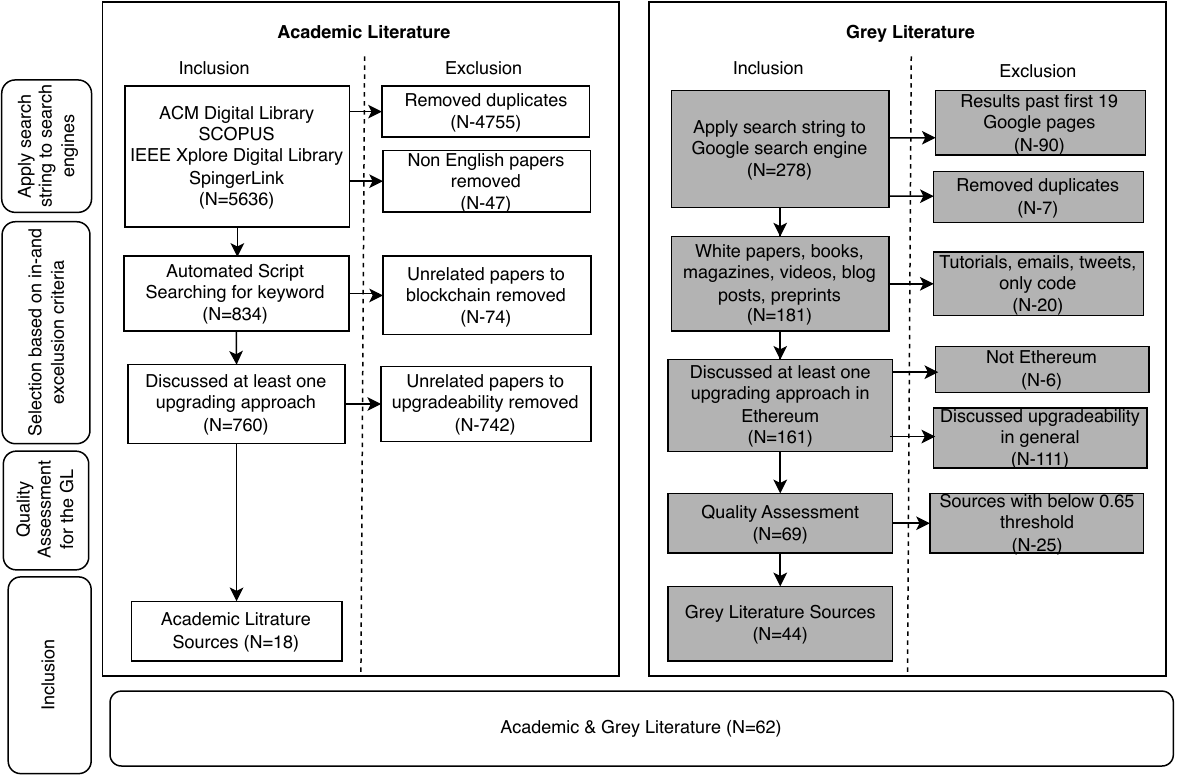}
    \caption{Overview of the source selection process for academic and grey literature}
    \label{fig:inclusion_exclusion}
\end{figure}

The Source Selection step outlines the approach taken to identify, include, and assess sources for this MLR, ensuring that only relevant academic and grey literature contributes to the analysis. Figure~\ref{fig:inclusion_exclusion} provides an overview of the source selection process applied.

\subsubsection{Grey Literature Sources}
Grey literature was an essential source for capturing real-world practices. Searches were conducted through Google to include a diverse range of sources, such as blogs, archived white papers, videos, and community-based forums. The search was conducted using Google in incognito mode to prevent personalization bias, and the first 10 pages of results (approximately 100 results) were reviewed. The search was extended incrementally until theoretical saturation was achieved, following the guidelines proposed by Garousi et al.~\cite{Garousi2019}. This ensured comprehensive coverage while maintaining search efficiency. Details of all collected grey literature sources from the initial search phase can be found in~\ref{app:A}.

\subsubsection{Academic Literature Sources}
To supplement the grey literature, academic databases including IEEE Xplore, SpringerLink, ACM Digital Library, and Scopus were searched. These databases were chosen for their extensive computer science and engineering coverage, particularly in blockchain and smart contract research. The refined search string was applied to each database, tailored to specific syntax requirements, and limited to English-language sources published from 2015 onwards (when Ethereum was introduced). To expand the search, backward and forward citation tracking was performed on key articles. The results of this search are detailed in~\ref{app:B}.

\subsubsection{Inclusion and Exclusion Process}
To ensure rigorous selection of sources, predefined inclusion and exclusion criteria were applied to both grey literature and academic sources. For academic literature, automated scripts identified sources by scanning titles, abstracts, and keywords for terms such as Blockchain, Ethereum, Smart Contract, and Solidity. This process ensured a targeted focus on relevant studies in the blockchain domain, minimizing irrelevant data. The criteria applied during the source selection process are summarized in Table~\ref{tab:inclusion_exclusion}. 

\begin{table}[t]
\caption{Inclusion and exclusion criteria applied for source selection}
\label{tab:inclusion_exclusion}
\resizebox{\textwidth}{!}{%
\begin{tabular}{cc}
\hline
\textbf{Inclusion Criteria} &
  \textbf{Exclusion Criteria} \\ \hline
\begin{tabular}[c]{@{}c@{}}Sources on smart contract upgrading \\ describing at least one pattern\end{tabular} &
  \begin{tabular}[c]{@{}c@{}}Non-English sources\\ Sources lacking technical depth \\ (e.g., brief news articles, opinion pieces without analysis)\\ Promotional content or advertisements\\ Duplicated GLs\\ Duplicate materials or summaries \\ without new insights\end{tabular} \\ \hline
\end{tabular}%
}
\end{table}

The screening procedure involved three independent reviewers who evaluated the titles, abstracts, and full texts of all identified sources. A majority rule decision approach was applied, where a source was included if at least two reviewers agreed on its relevance after the full-text review and excluded if at least two reviewers deemed it unsuitable. Discrepancies were resolved through discussion to ensure consistency in the selection process. To assess the reliability of this screening, Krippendorff’s Alpha was calculated, resulting in a score of 0.833, indicating high inter-rater reliability. Krippendorff’s Alpha measures inter-rater agreement, accounting for chance and is suitable for various data levels, with scores above 0.80 considered strong~\cite{Krippendorff2004}.~\ref{app:C} provides a detailed overview of the inclusion and exclusion decisions made during the selection process.

\subsubsection{Quality Assessment}
For grey literature, quality assessment was crucial, given the variability in source reliability. We adapted Garousi et al.'s quality assessment framework, covering 19 criteria grouped into seven quality categories: authority, methodology, objectivity, date, position with respect to related sources, novelty, impact, and outlet type. The first author independently assessed the quality of grey literature sources using the adapted framework. Each criterion was evaluated on a 3-point Likert scale (1 = Yes, 0.5 = Partly, 0 = No), and scores were averaged for a final quality score between 0 and 1.

After analyzing and calculating the quality scores for each source, it was essential to exclude low-quality sources to maintain the rigor and reliability of the review. In our case, we determined that any source with a score below 0.65 would be excluded. To establish this threshold, we employed an iterative quality threshold calibration process. We started with an initial high threshold of 0.8 and incrementally lowered it by 0.01, reviewing the sources that fell within each range. For each decrement, we assessed whether the sources added novel and valuable insights compared to those classified as high quality. This involved checking if the sources discussed broad patterns, provided limitations, or offered sufficient details that could be utilized, as opposed to merely listing information without enough depth. The process continued until we reached a threshold of 0.64, where sources below this score were consistently of low quality and did not add significant new details. To ensure thorough evaluation, sources with scores as low as 0.6 were also reviewed to gain a comprehensive understanding of quality distribution. This method ensured that the final threshold of 0.65 balanced inclusiveness with quality, aligning with the recommendations by Garousi et al.~\cite{Garousi2019} for maintaining robust criteria in MLRs. A detailed breakdown of the quality assessment for grey literature sources is provided in~\ref{app:D}.

\begin{table}[t]
\centering
\caption{Quality Categories and Criteria}
\label{tab:quality_categories}
\resizebox{\textwidth}{!}{%
\begin{tabular}{@{}lp{10cm}@{}}
    \toprule
    \textbf{Quality Category} & \textbf{Quality Criteria} \\
    \midrule
    Authority of the Producer & Is the publishing organization reputable? \\
                              & Is an individual author associated with a reputable organization? \\
                              & Has the author published other work in the field? \\
                              & Does the author have expertise in the area? \\
    \midrule
    Methodology               & Does the source have a clearly stated aim? \\
                              & Does the source have a stated methodology? \\
                              & Is the source supported by authoritative, contemporary references? \\
                              & Are any limits clearly stated? \\
                              & Does the work cover a specific question? \\
    \midrule
    Objectivity               & Is the statement in the source as objective as possible? \\
                              & Is there vested interest? \\
                              & Are the conclusions supported by data? \\
    \midrule
    Date                      & Does the item have a clearly stated date? \\
    \midrule
    Position w.r.t. Related Sources & Have key related GL or formal sources been linked to/discussed? \\
    \midrule
    Novelty                   & Does it enrich or add something unique to the research? \\
                              & Does it strengthen or refute a current position? \\
    \midrule
    Impact                    & Number of comments or views for specific online entries (blog posts, videos). \\
    \midrule
    Outlet Type               & 1st Tier GL (measure = 1) \\
                              & 2nd Tier GL (measure = 0.5) \\
                              & 3rd Tier GL (measure = 0) \\
    \bottomrule
\end{tabular}}
\end{table}

\subsubsection{Sources Statistics}
\begin{figure}[t]
    \centering
    \includegraphics[width=\textwidth]{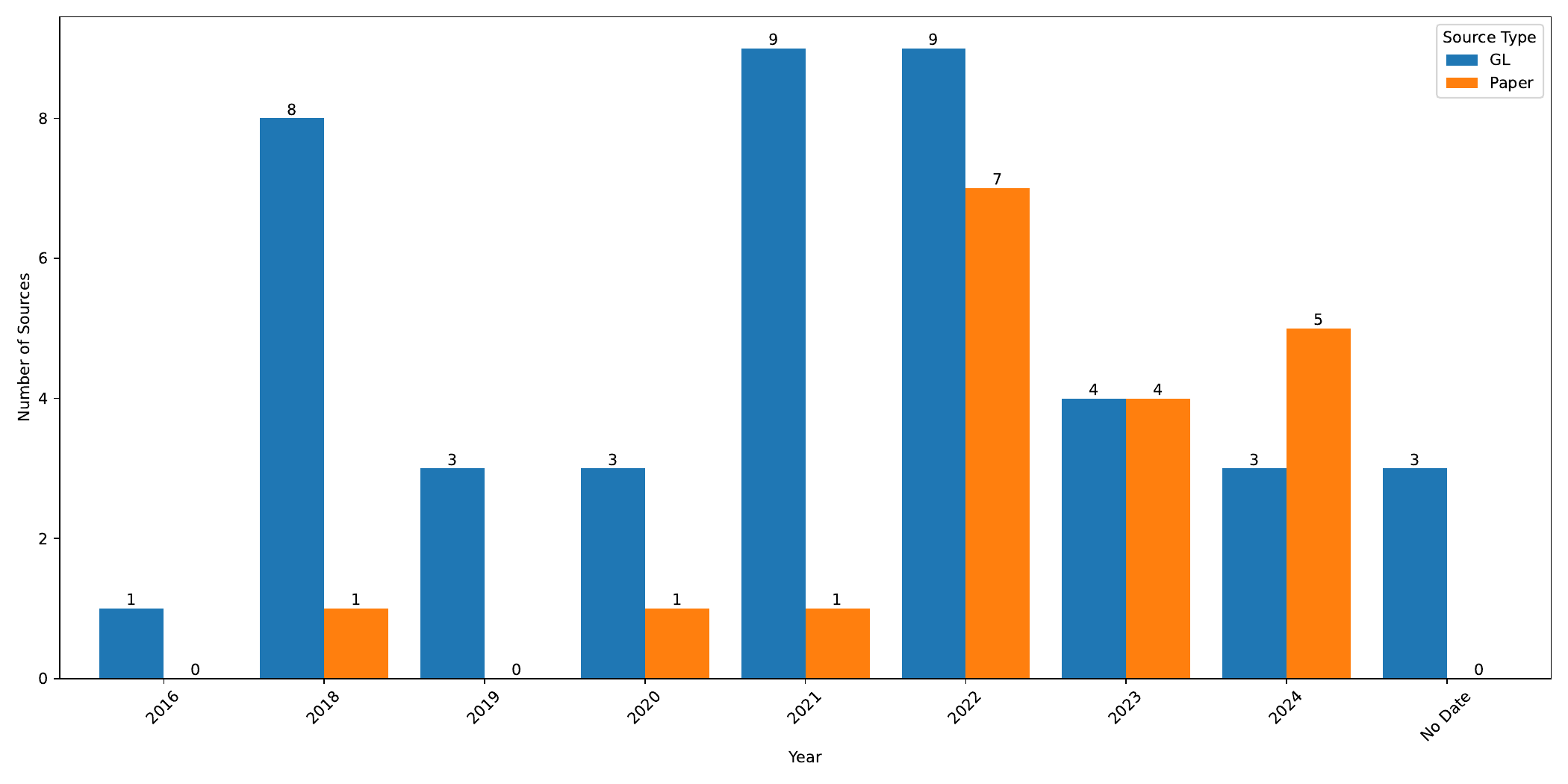}
    \caption{Distribution of included sources by publication year}
    \label{fig:mlr_statistics}
\end{figure}

A total of 62 sources were included in this study, comprising 19 academic papers and 43 grey literature (GL) sources, with data collected up to July 2024. The distribution of these sources by publication year is shown in Figure~\ref{fig:mlr_statistics}, where the x-axis represents the publication year and the y-axis indicates the number of sources. The figure highlights the increasing trend in publications over the years, demonstrating the growing interest and research in smart contract upgrade approaches. A comprehensive list of all included sources is provided in~\ref{app:E}.

\subsection{Data Extraction Process}
\subsubsection{Pilot Phase}
To refine the data extraction process, a pilot study was conducted with a selection of seven sources, which represented a mix of high and low-quality scores and included both grey and academic literature to test the extraction form's applicability across source types. A standardized data extraction form was developed to collect detailed descriptions of upgrade approaches, characteristics, and benefits or limitations linked to each approach, and notes to mention any extra information that does not fall in the predefined columns. Three reviewers independently extracted data from these sources and compared results to check consistency, as detailed in~\ref{app:F}.
Agreement rates were assessed using Krippendorff's Alpha, with the pilot showing high agreement for upgrade approaches (alpha = 0.948) and moderate agreement for characteristics and benefits/limitations (alpha = 0.761). Differences in interpretation were discussed, leading to revisions of the extraction form for clarity, including added definitions and clear definitions of each related aspect of the characteristics where the main differences were identified. The template used for data extraction is detailed in~\ref{app:G}.

\subsubsection{Data Extraction}
For the full data extraction, the same standardized form was used. Each reviewer independently extracted data from their assigned references, focusing on distinct sets of sources. To ensure consistency and address uncertainties, regular collaborative meetings were held throughout the extraction process. These discussions enabled reviewers to share insights, clarify ambiguities, and resolve uncertainties related to specific sources and approaches. A comprehensive summary of extracted data from all sources is provided in~\ref{app:H}.

\subsection{Data Synthesis Process}
In this study, we employed distinct data synthesis methodologies for each research question to systematically extract, analyze, and categorize the data from both academic and grey literature sources. These methodologies allowed us to provide a comprehensive answer to each research question while maintaining alignment with the structured approach of an MLR. Detailed descriptions of each method are provided in the corresponding sections and appendices.
\begin{enumerate}
    \item RQ1: Classification of Smart Contract Upgrade Approaches
    
    To classify upgrade approaches, we conducted Thematic Analysis~\cite{Braun2006}. This method allowed us to identify recurring concepts and themes in upgradeability, facilitating the classification into Full Upgrade and Partial Upgrade Approaches. Collaborative Consensus Classification sessions ensured robust categorization and unified naming conventions. Further details are available in Section~\ref{sec:RQ1}.
    \item RQ2: Characteristics of Upgrade Approaches

    For RQ2, we used Framework Analysis~\cite{Ritchie1994} to map each approach to its core components, address, logic, storage, and execution flow. This approach enabled us to categorize the upgrade methods based on characteristics such as address preservation, logic upgrade scope, storage management, and execution flow. Detailed descriptions can be found in Section~\ref{sec:RQ2}.
    \item RQ3: Benefits and Limitations of Upgrade Approaches

    We employed Thematic Analysis aligned with the ISO/IEC 25010 quality model to identify benefits and limitations. Each reviewer independently coded the benefits and limitations, followed by collaborative card-sorting sessions to group similar codes into broader themes such as complexity, flexibility, efficiency, security, and usability. Details of this process are provided in Section~\ref{sec:RQ3}.
\end{enumerate}

\section{Systematization of Knowledge}
\label{sec:systematization}
Before addressing the research questions, it is essential to systematize the key concepts surrounding upgradeability in smart contracts. Understanding the fundamental components of smart contracts is crucial, as these elements directly influence discussions on upgradeability and its implications.

\subsection{Fundamental Components of Smart Contracts}
\label{sec:Comp}
Smart contracts are composed of several fundamental components. The first is the \textbf{address}, a unique identifier on the blockchain through which users interact with the contract. The second is the \textbf{logic}, which defines the contract's functionality through its programmed rules, algorithms, and operations. The third is the \textbf{storage}, which maintains the contract's state, such as user balances or other persistent data. Lastly, the \textbf{execution flow} describes how transactions are processed, including the order of function calls and interactions within the contract.

These components are critical to understanding upgradeability because modifications to any of them can affect the contract's performance and the user experience. For instance, altering the logic may change the contract's functionality, while modifying storage can impact the integrity of the data maintained by the contract.

\subsection{A Systematic Definition of Smart Contract Upgradeability}
\label{sec:def}
In the literature, there are differing views on what constitutes upgradeability in smart contracts, reflecting a tension between the need for continuity and the desire for flexibility in post-deployment contract changes. Some researchers and practitioners define upgradeability as the ability to preserve the state of the contract across upgrades, ensuring minimal disruption for users while maintaining access to their data and balances~\cite{openzeppelin2020,chainlink2022, scsfg2023, hackernoon2020, bodell2023proxy, qasse2024immutable}. This perspective is particularly emphasized in contexts like decentralized finance (DeFi), where user continuity and trust are critical. Maintaining a seamless experience for users without requiring manual action during upgrades is often highlighted as a fundamental characteristic of an upgradeable smart contract.

Conversely, others consider approaches where state preservation is unnecessary for upgradeability~\cite{ethereum2024upgrading, salehi2022not,li2024characterizing}. In this view, deploying a new contract and requiring users to migrate data manually can still be classified as an upgradeable solution if it enables modifications to the contract logic, such as allowing new functionalities or significant updates. This interpretation broadens the scope of upgradeability to approaches that enable logical modifications, even if users must take extra steps to reconnect with the contract.

Additionally, perspectives differ on the level of flexibility necessary for upgradeability, particularly concerning the preservation of the interaction address. Some literature emphasizes approaches that allow modifications to any contract component while maintaining the original interaction address, ensuring seamless user interaction and adaptability for substantial system updates~\cite{hackernoon2018,logrocket2024uups,quicknode2024introduction}. Others discuss more open-ended approaches that do not maintain the original interaction address, allowing extensive changes to components and functionalities. This flexibility can result in users needing to interact with a new address after updates, thereby sacrificing consistency in user interaction~\cite{ethereum2024upgrading, salehi2022not,li2024characterizing}. While these more open approaches support significant modifications, they may impact user trust and continuity.

Given these differing views, there is a clear need to develop a unified definition of upgradeability that accommodates these perspectives while recognizing the unique constraints of smart contracts. Drawing from standard definitions of software upgradeability and considering the structure of smart contracts, we propose the following definition:

\textit{Upgradeability in smart contracts refers to the capability to modify the contract's logic or code post-deployment. This modification can occur with or without preserving the contract's state and may involve full or partial changes to the contract's components.}

This definition acknowledges that upgradeability can be achieved through various means, including both state-preserving and non-state-preserving approaches, and that it can involve modifications to different contract components. It emphasizes the importance of predefined mechanisms within the contract that allow for upgrades, distinguishing these from external interventions that may undermine the contract's autonomy.
\section{RO1: Smart Contract Upgrading Approaches}
\label{sec:RQ1}
To address RQ1, which aims to classify existing smart contract upgrade methods, we employed Thematic Analysis as our data extraction and synthesis method~\cite{Braun2006}. This approach allowed us to systematically identify, analyze, and report patterns within our collected data from both academic and grey literature sources. We began by thoroughly familiarizing ourselves with the literature to generate initial codes based on recurring concepts related to upgradeability. Guided by the unified definition of upgradeability established in Section~\ref{sec:def}, these codes were organized into potential themes, leading to the classification of upgrade methods into two primary categories: Full Upgrade Approaches and Partial Upgrade Approaches (discussed in detail in Section~\ref{RQ1-Class}).

This classification highlights the fundamental differences in how upgrades impact a contract's state and user interactions, providing a framework for evaluating existing approaches. To ensure the robustness of this classification, we conducted Collaborative Consensus Classification sessions among the three authors (Author 1, Author 2, and Author 3). During these discussions, we reviewed and deliberated on the identified approaches to reach an agreement on their categorization. Recognizing the varied terminologies used across sources, we also unified the naming conventions to standardize the terminology (see Table~\ref{tab:RQ1} in Section~\ref{RQ1:Un}). This table showcases the chosen unified names alongside the alternative names found in the literature, addressing potential ambiguities and improving clarity in communication. Through iterative refinement, we ensured that our classifications accurately represented the data and aligned with our unified definition of upgradeability. A detailed description of the Thematic Analysis process for RQ1 is provided in ~\ref{app:RQ1}.

\subsection{Classification of Smart Contract Upgrading Approaches}
\label{RQ1-Class}
The classification of smart contract upgrade methods is organized into two main categories: Full Upgrade Approaches and Partial Upgrade Approaches. This structure captures the differences in how each method modifies contract logic and storage, providing a clear framework for understanding the variety of upgrade techniques identified through thematic analysis.
\subsubsection{Full Upgrade Approaches}
In Full Upgrade approaches, the entire smart contract is redeployed when an upgrade is required. This process involves replacing both the contract's logic and state, resulting in a completely new contract version.

\begin{itemize}
    \item \textbf{Contract Migration:} Involves deploying a new smart contract at a different address to replace an existing one. The state, including balances and variables, is manually migrated from the old contract to the new one. Users must be informed of the change and update their interactions to the new contract address as the old contract becomes obsolete. This approach requires careful handling of state migration to avoid inconsistencies and may disrupt interactions with the system during the migration process. Developers often make official announcements to facilitate the transition, and platforms like Etherscan may label the old contract as "Old Contract" to notify users.
    
    \item \textbf{Metamorphic Contracts (CREATE2):} This upgrade mechanism leverages the CREATE2 opcode to redeploy a contract at the same address after the original contract self-destructs. Developers use \texttt{SELF\break DESTRUCT} to remove the existing contract's code and state, freeing up the address. A new contract with updated logic is then deployed to the same address using CREATE2, which allows pre-determination of the contract's address based on a hash of the deploying account, salt, and initialization code. While the contract's address remains the same, the state is not preserved, as using \texttt{SELFDESTRUCT} clears the state.
\end{itemize}

\subsubsection{Partial Upgrade Approaches}
Partial Upgrade approaches are designed to modify the smart contract's logic while preserving the contract's state. The two primary subcategories of partial upgrades are Two-Module Approaches and Multi-Module Approaches.

\paragraph{Two-Module Approaches}
Two-module approaches separate a contract's logic and state into two distinct modules, which allows for the full upgrade of the logic component through new deployments while the state remains stable. Two primary techniques within this category include Data Separation and Proxy-Based Approaches.

\begin{itemize}
    \item \textbf{Data Separation:} In this approach, a contract is divided into a logic contract that manages operations and a data contract that preserves the contract's state. Users interact with the logic contract, which references the data contract using opcodes like \texttt{CALL} or \texttt{STATICCALL} to access or modify data. To secure the state, the data contract restricts updates to those initiated by the designated logic contract address.

    For upgrades, developers deploy a new logic contract (at a new address) and update this address within the data contract's storage. This change requires users to interact with the new address, so developers often employ a registry that maintains the latest logic contract address. This allows users to retrieve the current address and interact seamlessly with the latest contract version. There are three main types of data separation: Inherited Storage, Eternal Storage, and Unstructured Storage.

    \begin{itemize}
        \item \textbf{Inherited Storage:} Maintains a consistent storage layout across upgrades using Solidity's inheritance mechanism. A separate Storage Contract defines all the state variables used by the contract. The logic contract and any upgraded versions inherit from this Storage Contract, ensuring they share the same storage structure. Developers must ensure the storage layout remains consistent, avoiding changes to existing state variables to prevent misalignment.

        \item \textbf{Eternal Storage:} Abstracts data storage into a separate contract using key-value pairs, typically implemented through mappings. The logic contract interacts with this storage contract using setters and getters to read or modify data. This allows developers to modify the logic contract freely without worrying about storage layout or data corruption.

        \item \textbf{Unstructured Storage:} Organizes data within specific storage slots identified by unique hashes, such as those generated by \texttt{keccak256}. The logic contract directly accesses and modifies these slots, often using low-level assembly code. This method eliminates the need for a predefined storage layout, optimizing data access efficiency.
    \end{itemize}

    \item \textbf{Proxy-Based Approaches:} Use a proxy contract to delegate calls to a logic contract while maintaining state within the proxy.
    
    \begin{itemize}
        \item \textbf{Basic Proxy:} The simplest form of the proxy pattern, holding state, and forwarding function calls to the logic contract using \texttt{delegatecall}. Users interact with the proxy, which maintains a consistent address. Upgrades involve deploying a new logic contract and updating the proxy's reference.

        \item \textbf{EIP-897 (Delegate Proxy):} Distinguishes between different proxy use cases, defining strict and upgradable proxies. It introduces interfaces that signal the proxy's purpose and primary implementation.

        \item \textbf{EIP-1967 (Standard Storage Slots):} Defines storage slots for the logic contract's address and upgrade-related variables. Upgrades involve updating the storage slot pointing to the logic contract.

        \item \textbf{UUPS (Universal Upgradeable Proxy Standard):} The proxy contract delegates calls to the logic contract but does not contain the upgrade mechanism. The logic contract includes the upgrade functionality.

        \item \textbf{Transparent Proxy:} Differentiates between admin and user functions, handling admin-only functions directly while forwarding user calls to the logic contract.

        \item \textbf{Beacon Proxy:} Uses a beacon contract to store the logic contract's address. Proxies retrieve the address from the beacon, allowing for simultaneous upgrades of multiple proxies.
    \end{itemize}
\end{itemize}

\paragraph{Multi-Module Approaches}
Multi-module approaches divide the contract's logic into multiple modules, each upgradable independently.

\begin{itemize}
    \item \textbf{Strategy Pattern:} Delegates logic to external strategy contracts, allowing targeted upgrades without altering the main contract.

    \item \textbf{EIP-1538 (Transparent Contract Standard):} Allows dynamic proxy-based upgrades, adding, replacing, or removing functions at runtime.

    \item \textbf{Diamond Pattern (EIP-2535):} Divides the contract into facets, each handling specific functions, enabling modular upgrades.
\end{itemize}

\paragraph{Hybrid Approaches}
Combine elements of data separation and proxy mechanisms to ensure storage compatibility.

\begin{itemize}
    \item \textbf{Proxy with Inherited Storage:} Aligns with Inherited Storage, where the logic contract inherits storage layout from the proxy.

    \item \textbf{Proxy with Eternal Storage:} Uses a separate storage contract, allowing logic contract upgrades without affecting the state.

    \item \textbf{Proxy with 
    Unstructured Storage:} Utilizes predefined storage slots for variables the logic contract manages.
\end{itemize}

\subsection{Unified Smart Contract Upgrading Terminologies}
\label{RQ1:Un}
Throughout our literature review, we encountered various terms used across sources to describe the same upgrade approaches, often creating ambiguity and making clear communication challenging. For instance, what we refer to as "Basic Proxy" is frequently labeled as "Proxy Contract," a term broadly applied to several proxy types. Similarly, "Data Segregation" is used variably, sometimes encompassing both what we classify as Data Separation and the Strategy Pattern, adding further confusion. Additionally, the term "Transparent" is inconsistently used to describe both EIP-1538 and Transparent Proxy approaches. 
We have adopted a unified terminology throughout our classification to address these inconsistencies for consistency and clarity. In~\ref{app:I}, we provide a comprehensive table mapping various names in the literature to our chosen standardized terms. This table serves as a reference, allowing readers to cross-reference and understand the relationships between different terms used in the field. Table~\ref{tab:RQ1} shows a sample of this table.

\begin{table}[t]
\centering
\caption{Standard Terminologies for Upgrading Approaches}
\label{tab:RQ1}
\resizebox{\textwidth}{!}{%
\begin{tabular}{ll}
\toprule
\textbf{Unified Term} & \textbf{Alternative Names} \\
\midrule
\textbf{Contract Migration} & Basic Contract Upgrade, Social Migration \\
\textbf{Data Separation} & Data separation pattern, Separate Logic Contract, Basic Data Segregation \\
\textbf{Basic Proxy} & Delegatecall-based proxies pattern, Proxy contract \\
\textbf{Strategy Pattern} & Updates through functions, Partially Upgradeable Contract Systems \\
\bottomrule
\end{tabular}%
}
\end{table}
\subsection{Statistics on Upgradeability Approaches}

\begin{figure}[t]
    \centering
    \includegraphics[width=\textwidth]{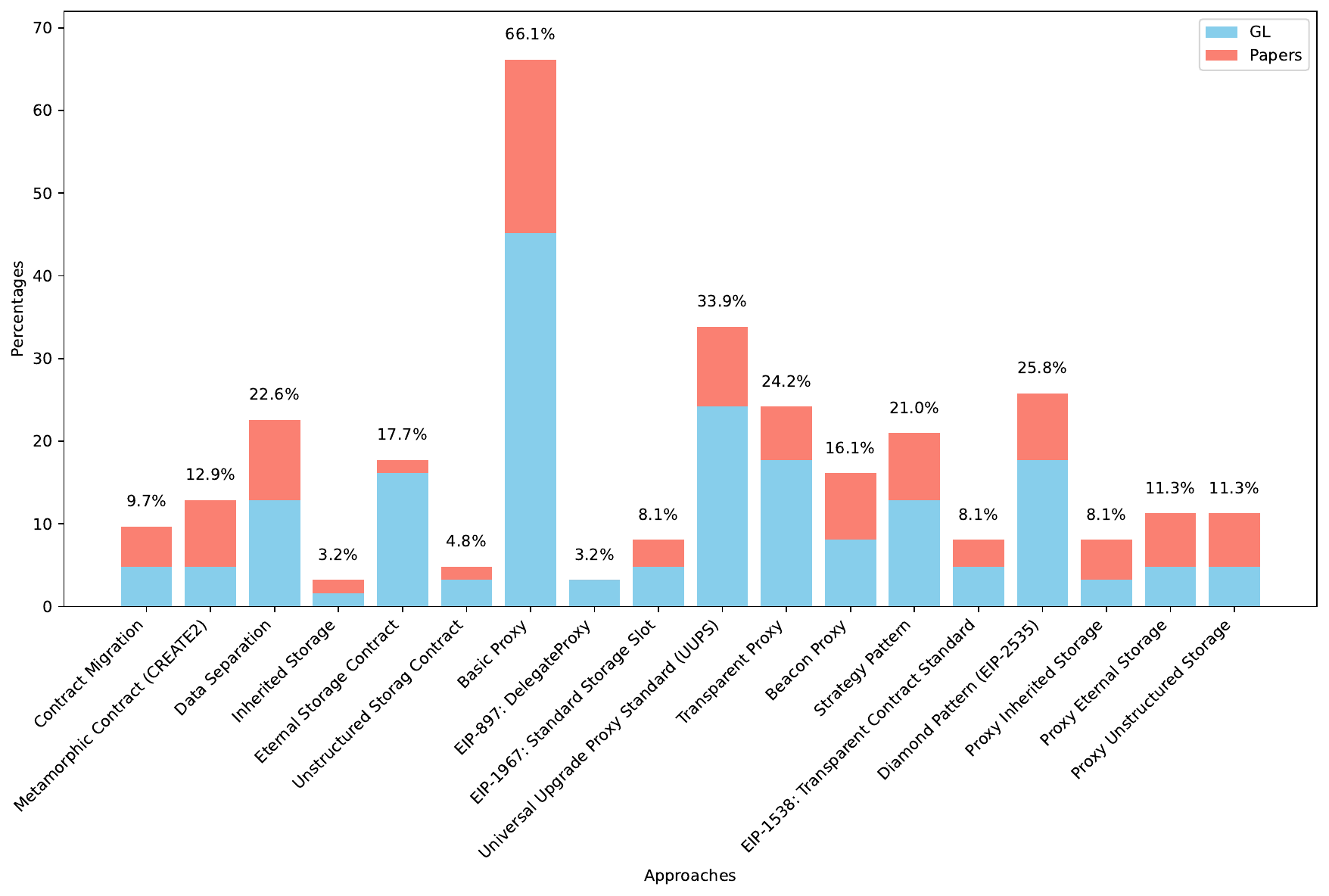} 
    \caption{Percentage of mentions for smart contract upgrade approaches across grey literature (GL) and academic papers}
    \label{fig:RQ1}
\end{figure}

Figure~\ref{fig:RQ1} presents a stacked bar chart depicting the distribution of mentions for different smart contract upgrade approaches across grey literature (GL) and academic papers, highlighting which techniques are most commonly discussed in each source type. Results show that Basic Proxy is the most frequently mentioned approach, aligning with the broadly accepted definition of upgradeability, which emphasizes preserving contract state across upgrades. Although other approaches, such as Data Separation, also maintain state continuity, proxies, especially Basic Proxy, dominate the conversation. Interestingly, a few studies (16.13\%) mention the Registry Pattern as an upgradeable mechanism; however, we do not consider it an upgrade approach per our definition, as it merely stores the latest contract version and can be used with other approaches. Additionally, proxy variations receive broad coverage in grey literature, while academic papers show more selective emphasis, especially on hybrid patterns such as Proxy with Eternal Storage. Lastly, data separation techniques, such as Eternal Storage, are largely discussed within grey literature, with only one academic mention~\cite{lopez2022upgradeable}, illustrating differing emphases across literature types.

\subsection{RQ1 Answer}
\begin{figure}[t!]
    \centering
    \includegraphics[width=\textwidth]{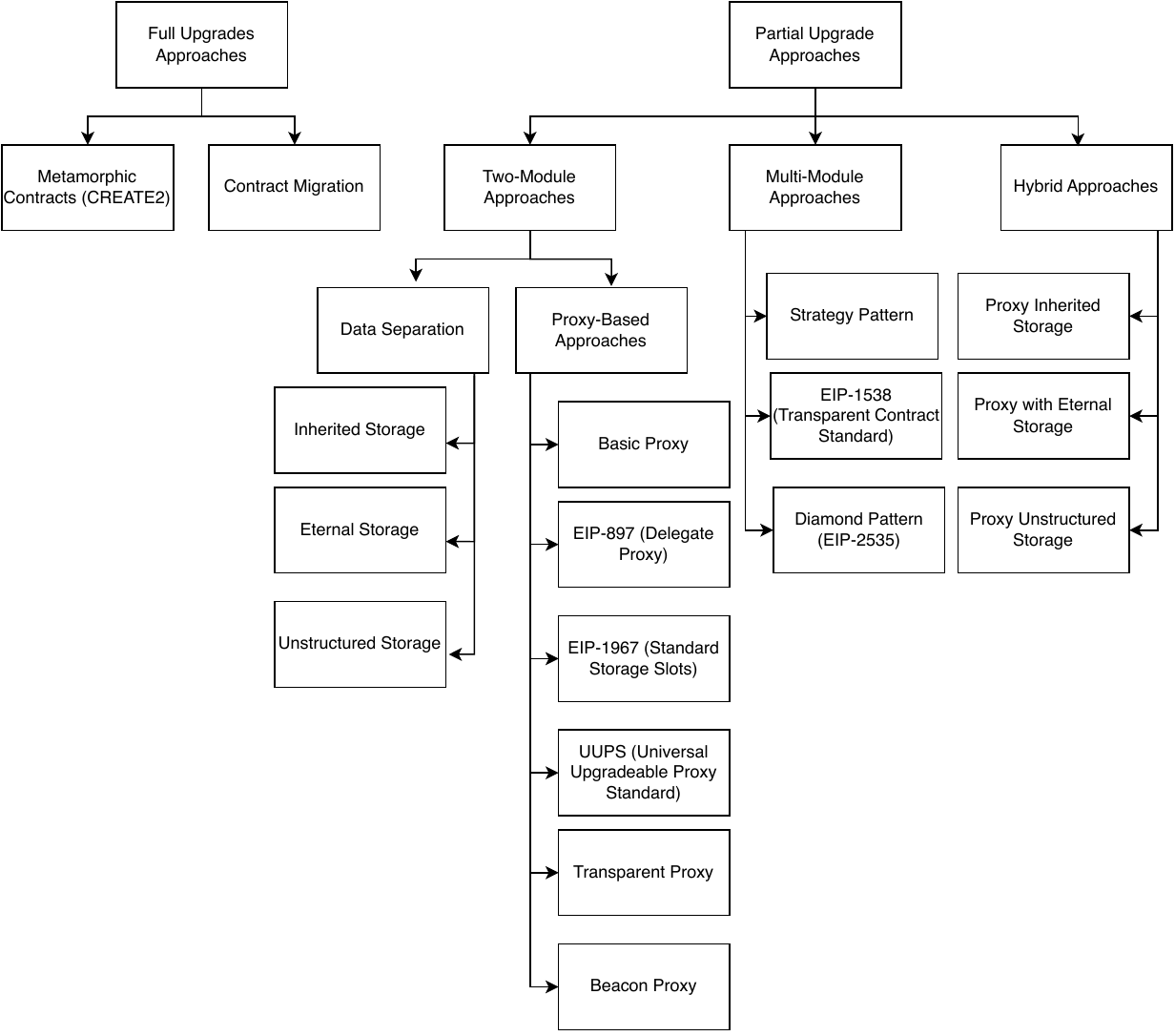}
    \caption{Overview of the classification of smart contract upgrade approaches}
    \label{fig:classification}
\end{figure}

To address \textbf{RQ1: What are the existing approaches for upgrading Ethereum smart contracts?}, we conducted a thematic analysis, identifying 17 distinct upgrade approaches. The classification of smart contract upgrades is structured around the scope of modifications(logic and storage). Figure~\ref{fig:classification} provides an overview of the proposed classification, dividing upgrade methods into two main categories, Full Upgrade Approaches, and Partial Upgrade Approaches, and detailing the specific methods identified through thematic analysis.
\section{RO2: Characteristics of Smart Contract Upgrading Approaches}
\label{sec:RQ2}
In this section, we address RQ2 by discussing and classifying the characteristics of various smart contract upgrade approaches, defined in RQ1, based on the core components of smart contracts introduced earlier (Section~\ref{sec:Comp}). To systematically analyze how each upgrade approach interacts with these components (address, logic, storage, and execution flow), we employed Framework Analysis as our data synthesis method~\cite{Ritchie1994}. This method is particularly suitable for applied research with specific questions and predefined issues, allowing us to organize and interpret data within a structured framework.

We began by mapping each upgrade approach to the core components, examining the specific ways in which they affect or utilize each component. This process involved familiarization with the data, identifying key themes related to each component, and indexing the data accordingly. We then charted the data into a matrix format, facilitating a clear comparison across different approaches.

By using Framework Analysis, we were able to categorize the upgrade approaches based on characteristics such as address preservation, logic upgrade scope, storage management, and execution flow. A detailed explanation of the Framework Analysis process for RQ2 is provided in~\ref{app:RQ2}.

\subsection{Address Preservation}
Address preservation determines whether the contract's address, which users interact with, remains consistent during an upgrade. This component has significant implications for user experience and interaction continuity. There are two types:
\begin{itemize}
    \item Fixed Address: These approaches ensure that users continue interacting with the same contract address even after an upgrade, enhancing trust and reducing the need for changes in user behavior. Proxy-based approaches (Basic Proxy, UUPS, EIP-1967, Transparent Proxy, Beacon Proxy), Multi-Module Approaches (Diamond Pattern, EIP-1538, Strategy Pattern), and Hybrid Approaches (Proxy Inherited Storage, Proxy with Eternal Storage, Proxy Unstructured Storage) fall under this type. Metamorphic Contracts (CREATE2) also maintain a fixed address, providing predictable user interactions.
    \item Changing Address: In these approaches, users interact with a new address every time there is an upgrade, leading to a fragmented experience and the need for users to update their references. This is characteristic of Contract Migration and Data Separation approaches (Inherited Storage, Eternal Storage, Unstructured Storage), where a new version of the contract is deployed at a different address.
\end{itemize}

\subsection{Logic Upgrade Scope}
The scope of a logic upgrade defines how extensively the logic of a contract can be modified during an upgrade, which impacts the system's flexibility. There are two types:
\begin{itemize}
    \item Full Logic Replacement: Approaches that allow complete replacement of the contract's logic to enable substantial changes. This is common in Full Upgrade Approaches (Contract Migration, Metamorphic Contracts), Proxy-Based Approaches (Basic Proxy, UUPS, EIP-1967, Transparent Proxy, Beacon Proxy), Hybrid Approaches (Proxy Inherited Storage, Proxy with Eternal Storage, Proxy Unstructured Storage), and Two-Module Approaches like Data Separation (Inherited Storage, Eternal Storage, Unstructured Storage).
    \item Partial Logic Replacement: Approaches that support updating specific logic modules without changing the entire contract. This type is seen in Multi-Module Approaches (Diamond Pattern, EIP-1538, Strategy Pattern), which offer targeted upgrades and modularity.
\end{itemize}
\subsection{Storage Management}
Storage management addresses how a contract's state is maintained during upgrades, which affects data integrity and continuity. This component has two types:
\begin{itemize}
    \item State Preservation: These approaches keep the contract's state intact during upgrades, maintaining data continuity and minimizing disruption. All approaches in the partial upgrade category in RQ1 (Two-Modules, Multi-Module, and Hybrid approaches) fall into this category. 
    \item State Migration Required: In these approaches, the contract's state must be transferred when a new version is deployed, introducing complexities and potential risks. Contract Migration and Metamorphic Contracts (CREATE2) typically require state migration.
\end{itemize}

\subsection{Execution Flow}
Execution flow refers to how function calls are processed within the contract and whether they are direct or delegated. This characteristic affects the modularity and performance of contract upgrades. There are two types:
\begin{itemize}
    \item Delegate Calls: Approaches that use delegatecall to execute logic in the context of the proxy's storage, facilitating logic upgrades while keeping the state consistent. Proxy-based approaches (Basic Proxy, UUPS, EIP-1967, Transparent Proxy, Beacon Proxy), Multi-Module Approaches, except Strategy pattern, and Hybrid Approaches use delegate calls. 
    \item Direct Calls: Approaches that execute function calls directly in the contract, simplifying the process but limiting modularity. Contract Migration, Metamorphic Contracts (CREATE2), Data separation approaches, and the Strategy Pattern within Multi-Module Approaches use direct calls.
\end{itemize}
\subsection{RQ2 Answer}
To answer RQ2: What are the characteristics of each upgrading approach?, we conducted a framework analysis of the approaches identified in RQ1. Table~\ref{tab:characteristics} provides an outline of the characteristics of smart contract upgrade approaches, highlighting how each approach manages address preservation, logic upgrade scope, storage management, and execution flow.  
\begin{table}[t!]
\centering
\caption{Characteristics of Smart Contract Upgrade Approaches Based on Core Components}
\label{tab:characteristics}
\resizebox{\textwidth}{!}{%
\begin{tabular}{lcccc}
\toprule
\textbf{Approach} & \textbf{Address Preservation} & \textbf{Logic Upgrade Scope} & \textbf{Storage Management} & \textbf{Execution Flow} \\
\midrule
Contract Migration & Changing Address & Full Replacement & State Migration Required & Direct Calls \\
Metamorphic Contracts (CREATE2) & Fixed Address & Full Replacement & State Migration Required & Direct Calls \\
Inherited Storage & Changing Address & Full Replacement & State Preservation & Delegate Calls \\
Eternal Storage & Changing Address & Full Replacement & State Preservation & Delegate Calls \\
Unstructured Storage & Changing Address & Full Replacement & State Preservation & Delegate Calls \\
Basic Proxy & Fixed Address & Full Replacement & State Preservation & Delegate Calls \\
UUPS & Fixed Address & Full Replacement & State Preservation & Delegate Calls \\
EIP-1967 & Fixed Address & Full Replacement & State Preservation & Delegate Calls \\
Transparent Proxy & Fixed Address & Full Replacement & State Preservation & Delegate Calls \\
Beacon Proxy & Fixed Address & Full Replacement & State Preservation & Delegate Calls \\
Diamond Pattern (EIP-2535) & Fixed Address & Partial Replacement & State Preservation & Delegate Calls \\
EIP-1538 & Fixed Address & Partial Replacement & State Preservation & Delegate Calls \\
Strategy Pattern & Fixed Address & Partial Replacement & State Preservation & Direct Calls \\
Proxy Inherited Storage & Fixed Address & Full Replacement & State Preservation & Delegate Calls \\
Proxy with Eternal Storage & Fixed Address & Full Replacement & State Preservation & Delegate Calls \\
Proxy Unstructured Storage & Fixed Address & Full Replacement & State Preservation & Delegate Calls \\
\bottomrule
\end{tabular}%
}
\end{table}

\section{RO3: Evaluation of Smart Contract Upgrade Approaches}
\label{sec:RQ3}
To systematically identify and categorize the benefits and limitations of each smart contract upgrade approach, we conducted a thematic analysis as part of our Multivocal Literature Review (MLR). Each reviewer independently coded the data, highlighting key benefits and limitations mentioned in the literature.

We then held collaborative sessions to discuss our findings, using card-sorting techniques to group similar codes into broader themes. This process allowed us to identify common evaluation criteria such as Complexity, Flexibility, Efficiency, Security, and Usability. We aligned these themes with recognized software quality models, particularly the ISO/IEC 25010 standard, adapting them to the specific context of smart contracts. In the following sections, we evaluate each upgrade approach against these metrics, considering the components above and comparing their benefits and limitations. Details of the thematic analysis process for RQ3 are provided in ~\ref{app:RQ3}.

\subsection{Complexity}

Complexity is crucial in evaluating smart contract upgrade approaches, as it impacts the effort needed to modify, maintain, or upgrade the system. According to ISO/IEC 25010, complexity, which corresponds to maintainability, is defined as:

\textit{The degree of effectiveness and efficiency with which the intended maintainers can modify a product or system.}

We assess complexity across three key aspects of smart contract components: logic, storage, and execution flow. We did not include the address component, as it does not influence the complexity.

\subsubsection{Logic Complexity}

Logic complexity refers to the challenge of updating or modifying a contract's logic without introducing errors. We classify logic complexity into four levels based on difficulty, where Low indicates ease of modification and Very High reflects significant difficulty.

\begin{itemize}
    \item \textbf{Low Logic Complexity:}
    \begin{itemize}
        \item \textit{Contract Migration:} Allows developers to rewrite the entire contract logic without needing to consider compatibility with previous versions, making the process straightforward and manageable~\cite{ethereum2024upgrading,hackernoon2020,salehi2022not,salehi2022analysis}.
        \item \textit{Strategy Pattern:} Enables isolated updates to specific functionalities without affecting the core contract, reducing the overall complexity of logic modifications~\cite{ebrahimiupc,li2024characterizing,openzeppelin2021security,salehi2022not}.
    \end{itemize}
    
    \item \textbf{Moderate Logic Complexity:}
    \begin{itemize}
        \item \textit{Data Separation Patterns (Inherited Storage, Eternal Storage, Unstructured Storage):} Maintain separate logic and storage, simplifying upgrades but requiring some coordination to ensure integration, resulting in moderate complexity~\cite{ethereum2024upgrading,medium2024proxy,trailofbits2018,indorse2018,salehi2022not}.
        \item \textit{UUPS Proxy:} Embeds the upgrade function directly within the logic contract, streamlining upgrades, though it requires caution to avoid issues like bricking~\cite{openzeppelin2020,moralis2022}.
        \item \textit{Metamorphic Contracts (CREATE2):} Support full rewrites of contract logic without complex structures, but redeployment steps demand attention to maintain the desired setup~\cite{openzeppelin2020,salehi2022not,wiesner2021,ethereumdeveloper2021upgrade,qasse2023smart}.
    \end{itemize}
    
    \item \textbf{High Logic Complexity:}
    \begin{itemize}
        \item \textit{Proxy-Based Approaches (Basic Proxy, Transparent Proxy, Beacon Proxy, EIP-897, EIP-1967):} Rely heavily on \texttt{delegatecall}, which increases complexity due to the need for precise function selector management and consistent mapping to avoid clashes~\cite{trailofbits2018,jeiwan2021,soliditydeveloper2024, du2023four,qasse2024immutable}.
        \item \textit{Hybrid Approaches (Proxy with Eternal Storage, Proxy Inherited Storage, Proxy Unstructured Storage):} Combine proxy functionality with specific storage handling, requiring careful control of storage pointers and proxy forwarding, adding significant complexity~\cite{ebrahimiupc,klinger2020upgradeability, bui2021evaluating, openzeppelin2020}.
    \end{itemize}
    
    \item \textbf{Very High Logic Complexity:}
    \begin{itemize}
        \item \textit{Diamond Pattern (EIP-2535):} The facet-based structure allows extensive modularity but requires precise management of function selectors and alignment across facets to avoid misalignment and unintended interactions~\cite{ethereum2024upgrading,openzeppelin2020,qasse2024immutable, openzeppelin2020,logrocket2024uups, wiesner2021,blockchains2024,eip2535diamonds2022}.
        \item \textit{EIP-1538 (Transparent Contract Standard):} Supports dynamic addition and removal of functions, which demands ongoing adjustments in function mapping and careful oversight to prevent clashes and logic errors~\cite{wiesner2021, ethereumdeveloper2021upgrade,eip1538}.
    \end{itemize}
\end{itemize}

\subsubsection{Storage Complexity}

Storage Complexity refers to the effort to manage or maintain the contract's state during upgrades. We assess storage complexity based on difficulty, where Low indicates straightforward state management, and Very High signifies significant challenges.
\begin{itemize}
    \item \textbf{Low Storage Complexity:}
    \begin{itemize}
        \item \textit{Proxy-Based Approaches:} This includes Basic Proxy, Transparent Proxy, Beacon Proxy, EIP-897, UUPS Proxy, and EIP-1967 Standard Storage Slots. These approaches maintain state within the proxy itself, avoiding the need for state migration or extensive coordination, making storage management straightforward~\cite{trailofbits2018, blockmagnates2022upgradability,salehi2022not,moralis2022}.
        \item \textit{Strategy Pattern:} Isolates state within the core contract, with individual strategies operating independently without affecting core storage. This simplicity results in a low level of storage complexity~\cite{ebrahimiupc, li2024characterizing, openzeppelin2021security, salehi2022not,qasse2023smart}.
    \end{itemize}
    
    \item \textbf{Moderate Storage Complexity:}
    \begin{itemize}
        \item \textit{Data Separation Patterns:}
        \begin{itemize}
            \item \textit{Unstructured Storage:} Flexible mappings and hash-based keys simplify storage management without strict adherence to predefined structures~\cite{runtime2022,bodell2023proxy}.
            \item \textit{Inherited Storage:} Provides consistent storage layout across contracts, requiring careful but not overly complex alignment~\cite{ethereum2024upgrading, medium2024proxy,indorse2018,salehi2022not,ebrahimi2024large}.
        \end{itemize}
        \item \textit{Hybrid Approaches (Proxy with Inherited Storage, Proxy Unstructured Storage):} Combine proxy mechanisms with simpler storage contracts, adding moderate complexity through pointer management and ensuring storage consistency across contracts~\cite{ebrahimiupc,klinger2020upgradeability,bui2021evaluating}.
    \end{itemize}
    
    \item \textbf{High Storage Complexity:}
    \begin{itemize}
        \item \textit{Data Separation - Eternal Storage:} Involves complex storage access due to its abstract structure. Managing data types like structs and mappings increases the complexity of updates and alignment, making it more challenging than other Data Separation types~\cite{securityboulevard2018,indorse2018, flolio2022,lopez2022upgradeable}.
        \item \textit{Hybrid Approaches (Proxy with Eternal Storage):} Eternal Storage within a hybrid setup requires detailed management of storage mappings and state consistency, adding high storage complexity~\cite{ebrahimiupc,klinger2020upgradeability,bui2021evaluating}.
        \item \textit{Diamond Pattern (EIP-2535):} The modular structure requires shared storage among facets, necessitating careful layout management to prevent collisions and ensuring that facets interact correctly with shared storage~\cite{ethereum2024upgrading,openzeppelin2020,qasse2024immutable, openzeppelin2020,logrocket2024uups, wiesner2021,blockchains2024,eip2535diamonds2022}.
        \item \textit{EIP-1538 (Transparent Contract Standard):} Similar to Diamond, its dynamic function modifications require consistent storage updates across modules, adding to storage management demands~\cite{wiesner2021,ethereumdeveloper2021upgrade,eip1538,qasse2024immutable}.
    \end{itemize}
    
    \item \textbf{Very High Storage Complexity:}
    \begin{itemize}
        \item \textit{Contract Migration:} Involves full state migration to a new contract, requiring extensive manual data transfer and state alignment, resulting in significant resource and time costs~\cite{ethereum2024upgrading,hackernoon2020,salehi2022not,salehi2022analysis}.
        \item \textit{Metamorphic Contracts (CREATE2):} Lack state preservation post-self-destruction, necessitating external management for any retained data. Reinitializing the state upon redeployment leads to very high storage complexity for continuity~\cite{openzeppelin2020,salehi2022not, li2024characterizing, ethereumdeveloper2021upgrade,huang2024sword}.
    \end{itemize}
\end{itemize}

\subsubsection{Execution Flow Complexity}

Execution Flow Complexity refers to the complexity introduced in the contract's execution flow due to the upgrade mechanism. Similar to the previous components, we rank this complexity based on difficulty, where Low means minimal changes and Very High means major challenges.

\begin{itemize}
    \item \textbf{Low Execution Flow Complexity:}
    \begin{itemize}
        \item \textit{Contract Migration:} Simple execution flow with no complex delegation or routing requirements. Users interact directly with the new contract, and upgrades involve full redeployment rather than layered routing.
        \item \textit{Strategy Pattern:} The execution flow remains straightforward since each strategy operates independently, allowing the core contract to delegate calls to specific strategies without impacting the main execution structure~\cite{ethereum2024upgrading,hackernoon2020}.
        \item \textit{Metamorphic Contracts (CREATE2):} Execution remains simple as each upgrade involves redeployment rather than intricate call routing or storage dependencies~\cite{li2024characterizing,ethereumdeveloper2021upgrade, huang2024sword}.
    \end{itemize}
    
    \item \textbf{Moderate Execution Flow Complexity:}
    \begin{itemize}
        \item \textit{Proxy-Based Approaches (including Basic Proxy, UUPS Proxy, EIP-897, and EIP-1967):} These proxies add a moderate level of complexity by using \texttt{delegatecall} to route calls to the logic contract. However, they do not require additional role-based distinctions or external dependencies~\cite{trailofbits2018,lohr2020maintenance,certik2022,ethereum2024upgrading,qasse2023smart,qasse2024immutable}.
        \begin{itemize}
            \item \textit{Basic Proxy:} Manages straightforward \texttt{delegatecall} routing to the logic contract.
            \item \textit{UUPS Proxy:} Stores the upgrade function within the logic contract itself, requiring fewer admin roles but still necessitating \texttt{delegatecall} management during upgrades.
            \item \textit{EIP-897 and EIP-1967:} Implement similar \texttt{delegatecall} routing while avoiding complex role management.
        \end{itemize}
        \item \textit{Data Separation Patterns (Unstructured Storage, Inherited Storage):} While they involve an additional call to access separate storage, their execution flow remains relatively simple. The logic and storage layers are distinct, but no significant routing complexity is added beyond storage retrieval~\cite{ethereum2024upgrading,medium2024proxy,trailofbits2018,indorse2018,salehi2022not}.
    \end{itemize}
    
    \item \textbf{High Execution Flow Complexity:}
    \begin{itemize}
        \item \textit{Transparent Proxy:} Requires additional role management to separate admin from user functions, creating layers in the execution flow and increasing complexity, especially when managing permissions for upgrades~\cite{moralis2022, scsfg2023}.
        \item \textit{Beacon Proxy:} Adds an external dependency on a centralized beacon contract for logic contract addresses, adding a routing step and reliance on the beacon for directing proxies~\cite{amri2023review,wasnik2024proxy,ebrahimiupc}.
        \item \textit{Hybrid Approaches (Proxy with Eternal Storage, Proxy Inherited Storage, Proxy Unstructured Storage):} Combine proxy routing with specific storage contracts, introducing higher execution flow complexity due to the additional storage contract coordination and external calls~\cite{ebrahimiupc,klinger2020upgradeability,bui2021evaluating,mvpworkshop2021}.
    \end{itemize}
    
    \item \textbf{Very High Execution Flow Complexity:}
    \begin{itemize}
        \item \textit{Diamond Pattern (EIP-2535)} and \textit{EIP-1538 (Transparent Contract Standard):} These introduce very high execution flow complexity, as managing multiple facets or dynamically adding and removing functions increases both the routing complexity and the risk of execution path mismanagement~\cite{eip2535diamonds2022,eip1538,salehi2022not,wiesner2021}.
    \end{itemize}
\end{itemize}

\subsubsection{Summary of Complexity Rankings}

\begin{figure}[t]
    \centering
    \includegraphics[width=\textwidth]{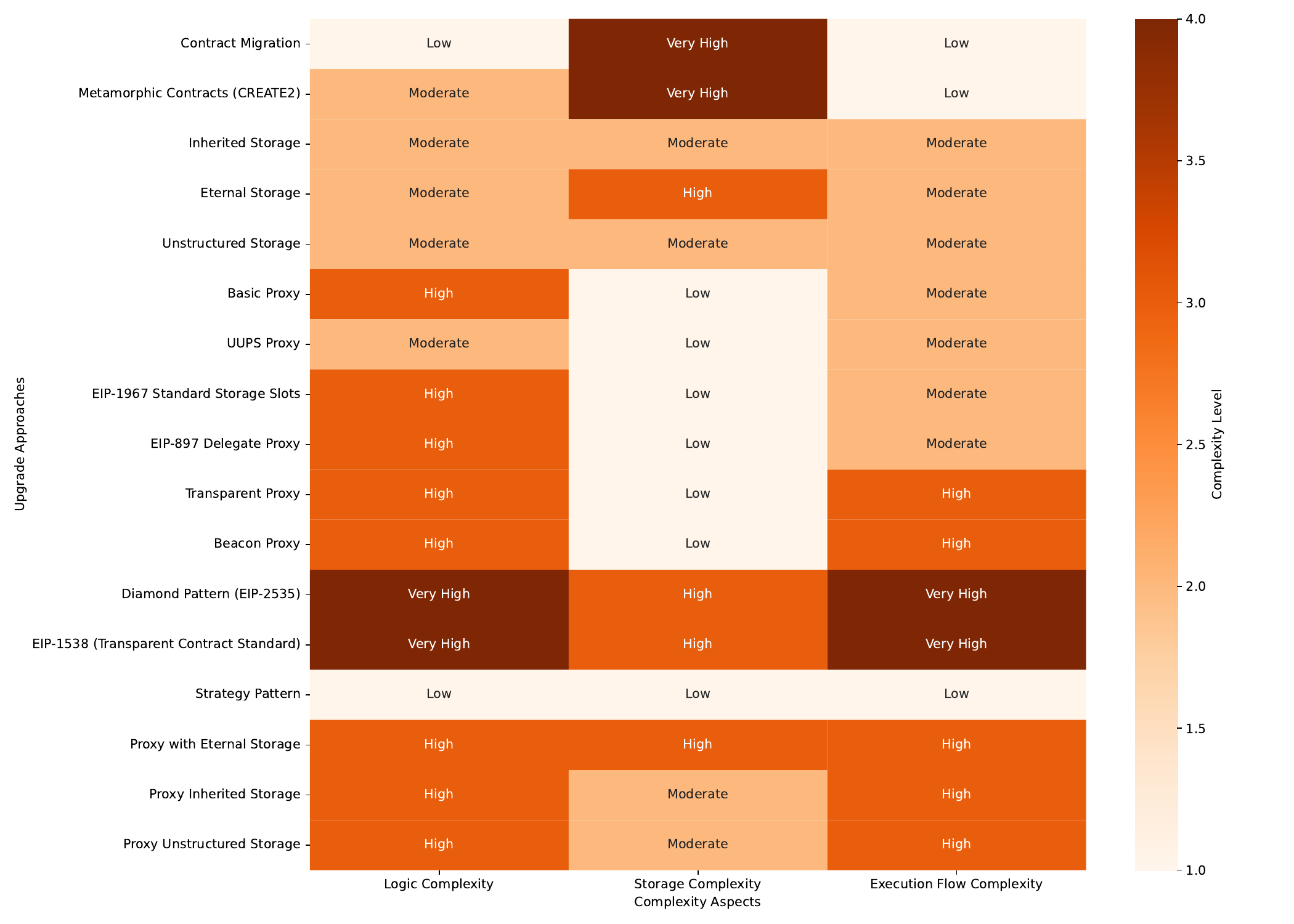} 
    \caption{Complexity levels for smart contract upgrade approaches}
    \label{fig:RQ3-comp}
\end{figure}

The heatmap, presented in Figure~\ref{fig:RQ3-comp}, provides a comprehensive visual summary of the complexity levels across various smart contract upgrade approaches. The X-axis represents the three aspects of complexity (Logic Complexity, Storage Complexity, and Execution Flow Complexity), while the Y-axis lists each smart contract upgrade approach individually. Color coding is utilized to indicate complexity levels, ranging from low to very high. Darker colors signify higher complexity, while lighter shades represent lower complexity.

We calculated an average complexity score for each approach across the three aspects to classify and interpret the complexity data. Complexity scores were assigned based on a 1-4 scale, where 1 is low complexity, and 4 is very high complexity. Using these scores, we derived an overall complexity classification for each approach as follows:

\begin{itemize}
    \item \textbf{Low Complexity:}
    \begin{itemize}
        \item \textit{Strategy Pattern}: Exhibits the lowest complexity among all approaches, scoring 1. Its isolated, core-focused structure enables upgrades to specific functionalities without affecting the primary execution or storage setup.
        \item \textit{Contract Migration}: Also falls into the low complexity category, allowing developers to rewrite the entire contract logic without considering compatibility with previous versions. This straightforward process makes it manageable and efficient.
    \end{itemize}
    
    \item \textbf{Moderate Complexity:}
    \begin{itemize}
        \item \textit{Proxy-Based Approaches}: Including Basic Proxy, Transparent Proxy, Beacon Proxy, UUPS Proxy, EIP-897, and EIP-1967, maintain moderate complexity levels. These approaches streamline upgrades but require careful function management due to the use of \texttt{delegatecall}, which adds layers of complexity to both the logic and execution flow.
        \item \textit{Data Separation Patterns}: Such as Inherited Storage, Eternal Storage, and Unstructured Storage exhibit moderate complexity as well. While these patterns allow for separate management of logic and storage, they still require coordination to ensure integration.
        \item \textit{Metamorphic Contracts (CREATE2)}: Also fit into this category due to their capacity for full logic rewrites, which simplifies some aspects of complexity but introduces challenges regarding state preservation.
    \end{itemize}
    
    \item \textbf{High Complexity:}
    \begin{itemize}
        \item \textit{Hybrid Approaches}: Which include Proxy with Eternal Storage, Proxy Inherited Storage, and Proxy Unstructured Storage, are classified as high complexity due to their combination of proxy functionalities with specific storage management. These approaches demand careful coordination between the proxy and storage layers, increasing the overall complexity.
    \end{itemize}
    
    \item \textbf{Very High Complexity:}
    \begin{itemize}
        \item \textit{Diamond Pattern (EIP-2535)} and \textit{EIP-1538 (Transparent Contract Standard)}: Fall into the very high complexity category. Both approaches involve extensive modularity and dynamic function management, necessitating careful management of function selectors and alignment across facets to avoid unintended interactions.
    \end{itemize}
\end{itemize}
\subsection{Flexibility}

Flexibility is an important factor in smart contract upgrade approaches, as it indicates how easily a contract can be adapted to meet changing requirements. According to ISO/IEC 25010, adaptability is defined as:

\textit{The degree to which a product or system can be effectively and efficiently adapted for different or evolving hardware, software, or other operational or usage environments.}

We evaluate flexibility in terms of logic and storage, as these components significantly impact how adaptable a contract is. Execution flow and address are not included in this evaluation because they do not directly influence the contract's adaptability.

\subsubsection{Logic Flexibility}

Logic Flexibility refers to how easily a smart contract's logic can be updated, extended, or modified to add new features or functionality without significant limitations. We rank logic flexibility in four levels, where low indicates limited adaptability and very high means extensive adaptability and ease of modification.

\begin{itemize}
    \item \textbf{Low Logic Flexibility:}
    \begin{itemize}
        \item \textit{Contract Migration}: Updating individual functions is not possible, as each change requires deploying a completely new contract, limiting flexibility to refine or expand existing logic incrementally~\cite{ethereum2024upgrading,hackernoon2020,salehi2022not,salehi2022analysis}.
        \item \textit{Metamorphic Contracts (CREATE2)}: Although redeployment is possible at the same address, the entire logic must be replaced each time, as isolated updates to specific functions are unsupported~\cite{wiesner2021,ebrahimiupc}.
    \end{itemize}
    
    \item \textbf{Moderate Logic Flexibility:}
    \begin{itemize}
        \item \textit{Inherited Storage}: New functions can be introduced through inheritance, but modifying inherited logic poses challenges, limiting adaptability for direct updates to existing functions~\cite{openzeppelin2020, medium2024proxy,indorse2018,bodell2023proxy}.
        \item \textit{Strategy Pattern}: Functionality within specific areas can be updated by swapping strategy contracts, allowing isolated changes without altering the primary logic, but broader updates are limited~\cite{bui2021evaluating,ethereum2024upgrading,yos2018}.
    \end{itemize}
    
    \item \textbf{High Logic Flexibility:}
    \begin{itemize}
        \item \textit{Data Separation Patterns (Eternal Storage, Unstructured Storage)}: With storage managed separately, new functions or modifications can be introduced to the logic without affecting stored data, supporting flexibility in evolving logic~\cite{securityboulevard2018,indorse2018, bodell2023proxy,runtime2022}.
        \item \textit{Proxy-Based Approaches (Basic Proxy, Transparent Proxy, Beacon Proxy, EIP-897, EIP-1967, UUPS)}: These proxies allow the replacement of the implementation contract, supporting the addition of new functions or modifications to existing ones within the constraints of the proxy structure~\cite{trailofbits2018,soliditydeveloper2024,openzeppelin2020,ethereum2024upgrading,salehi2022not,wasnik2024proxy}.
    \end{itemize}
    
    \item \textbf{Very High Logic Flexibility:}
    \begin{itemize}
        \item \textit{Diamond Pattern (EIP-2535)}: Facet-based structure enables modular updates, allowing functions to be added, replaced, or removed independently within facets, offering highly adaptable logic~\cite{bodell2023proxy, hacken2023,salehi2022not,eip2535diamonds2022,qasse2024immutable}.
        \item \textit{EIP-1538 (Transparent Contract Standard)}: Supports on-the-fly updates to add, replace, or remove functions dynamically, providing unrestricted flexibility for granular modifications to contract functionality~\cite{eip2535diamonds2022,eip1538,salehi2022not,wiesner2021,qasse2024immutable}.
    \end{itemize}
\end{itemize}

\subsubsection{Storage Flexibility}

Storage Flexibility refers to the ability to modify storage structures, such as adding new state variables or altering existing data without causing conflicts or inconsistencies. Similar to logic flexibility, we rank storage flexibility into four levels, where Low indicates limited adaptability and Very High represents extensive adaptability and ease of modification.

\begin{itemize}
    \item \textbf{Low Storage Flexibility:}
    \begin{itemize}
        \item \textit{Contract Migration}: Storage is reset during migration, and the layout is fixed; updates to existing data structures or the addition of new variables are restricted~\cite{hackernoon2020,li2024characterizing,salehi2022not}.
        \item \textit{Metamorphic Contracts (CREATE2)}: Each redeployment resets storage, and maintaining prior data is difficult, limiting flexibility to make incremental updates to the storage structure~\cite{openzeppelin2020,salehi2022not,li2024characterizing}.
        \item \textit{Inherited Storage and Proxy Inherited Storage}: Storage layout follows the inheritance hierarchy, and adding or modifying variables can disrupt the existing structure, reducing flexibility in adjusting stored data.
        \item \textit{Proxy-Based Approaches (Basic Proxy, Transparent Proxy, Beacon Proxy, EIP-897, EIP-1967, UUPS)}: The proxy structure requires compatibility across implementation changes, which restricts flexibility in modifying storage as it must follow a consistent layout~\cite{medium2024proxy,chainlink2022,wasnik2024proxy,bodell2023proxy}.
    \end{itemize}
    
    \item \textbf{Moderate Storage Flexibility:}
    \begin{itemize}
        \item \textit{Strategy Pattern}: While strategy contracts can change logic, the primary contract's storage structure remains largely static, offering limited flexibility for evolving storage needs~\cite{bui2021evaluating, ethereum2024upgrading,yos2018}.
        \item \textit{EIP-1538}: Functions can be adjusted dynamically, but adding or modifying state variables must be managed carefully to align with existing storage, offering controlled flexibility~\cite{eip2535diamonds2022,eip1538,salehi2022not,wiesner2021,qasse2024immutable}.
    \end{itemize}
    
    \item \textbf{High Storage Flexibility:}
    \begin{itemize}
        \item \textit{Eternal Storage}: With storage fully separated from logic, state variables can be added or modified independently, allowing flexible updates to stored data without impacting contract functionality~\cite{securityboulevard2018,indorse2018,bodell2023proxy}.
        \item \textit{Proxy Eternal Storage}: Storage is managed separately, which allows independent updates to state variables, providing high flexibility for adjusting storage as needed~\cite{openzeppelin2020,mvpworkshop2021,flolio2022}.
    \end{itemize}
    
    \item \textbf{Very High Storage Flexibility:}
    \begin{itemize}
        \item \textit{Diamond Pattern (EIP-2535)}: The layout structure prevents conflicts, allowing extensive modifications to stored data without risking disruption to other facets, enabling adaptable storage updates~\cite{ebrahimi2024large,soliditydeveloper2024,eip2535diamonds2022}.
        \item \textit{Unstructured Storage}: Unique storage slots for each variable allow new data to be added or modified without affecting existing storage, providing adaptability in adjusting the storage structure.
        \item \textit{Proxy Unstructured Storage}: With distinct slots for each variable, adding or modifying storage data is straightforward, achieving high adaptability without conflicts~\cite{mvpworkshop2021, flolio2022}.
    \end{itemize}
\end{itemize}

\subsubsection{Summary of Flexibility Rankings}

\begin{figure}[t]
    \centering
    \includegraphics[width=\textwidth]{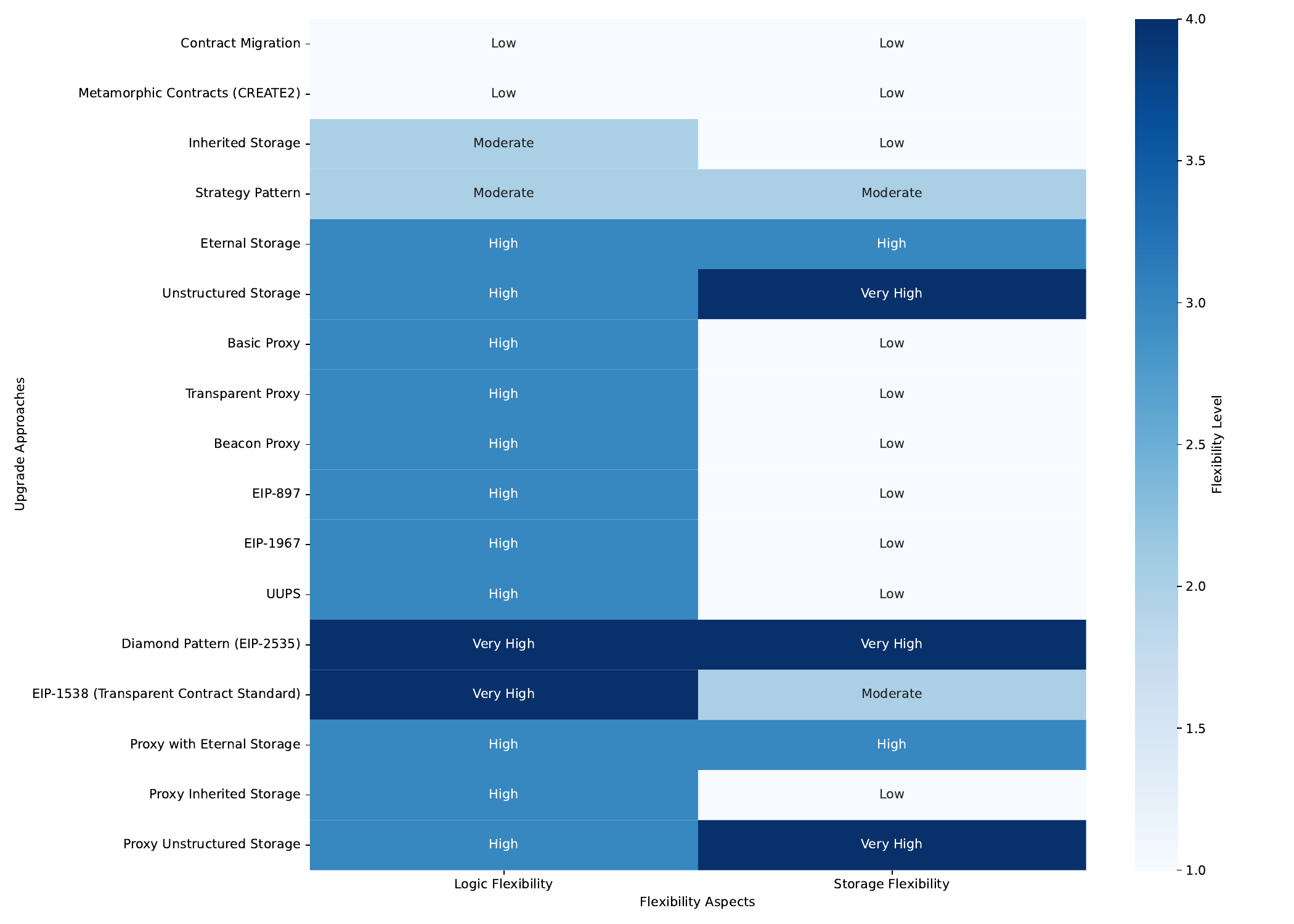} 
    \caption{Flexibility levels for smart contract upgrade approaches}
    \label{fig:RQ3-flex}
\end{figure}

As illustrated in Figure~\ref{fig:RQ3-flex}, the heatmap provides an integrated view of flexibility levels across a range of smart contract upgrade mechanisms. On the X-axis, we depict the two primary dimensions of flexibility (Logic Flexibility and Storage Flexibility), while the Y-axis lists each smart contract upgrade approach. A color gradient represents low to very high flexibility levels, where darker tones highlight higher flexibility and lighter tones signify lower flexibility.
We assigned an average flexibility score to each approach to systematically classify and interpret the flexibility data, encompassing both Logic and Storage Flexibility dimensions. Scores range from 1 to 4, with 1 indicating low flexibility and 4 representing very high flexibility. Based on these average scores, we categorized each approach's overall flexibility level as follows:

\begin{itemize}
    \item \textbf{Low Flexibility:}
    \begin{itemize}
        \item Approaches like \textit{Contract Migration} and \textit{Metamorphic Contracts (CREATE2)} exhibit low flexibility. Both require redeployment for any update, making isolated function modifications or incremental logic refinements impractical. Consequently, changes entail deploying entirely new contracts, resetting the state, and limiting adaptability.
    \end{itemize}
    
    \item \textbf{Moderate Flexibility:}
    \begin{itemize}
        \item The \textit{Strategy Pattern} and \textit{Inherited Storage} approaches fall under moderate flexibility. While the Strategy Pattern allows isolated changes through swappable strategy contracts, broader updates are constrained. Inherited Storage enables the introduction of new functions through inheritance but restricts modifications to inherited logic.
    \end{itemize}
    
    \item \textbf{High Flexibility:}
    \begin{itemize}
        \item \textit{Proxy-Based Approaches}, including Basic Proxy, Transparent Proxy, Beacon Proxy, EIP-897, EIP-1967, and UUPS, offer high logic flexibility, allowing implementation contract replacements. However, the requirement for consistent storage structures limits their adaptability in storage. \textit{Data Separation Patterns} like Eternal Storage and Unstructured Storage are similarly flexible, as they decouple storage from logic, allowing logic updates without affecting data.
    \end{itemize}
    
    \item \textbf{Very High Flexibility:}
    \begin{itemize}
        \item The \textit{Diamond Pattern (EIP-2535)} and \textit{EIP-1538 (Transparent Contract Standard)} approaches are classified as very high flexibility. With extensive modularity and dynamic function management, these approaches permit granular modifications and unrestricted adaptability in both logic and storage.
    \end{itemize}
\end{itemize}

\subsection{Efficiency}

Efficiency is a critical factor in smart contract upgradeability, as it directly impacts the cost and performance of contract operations on decentralized systems. According to ISO/IEC 25010, performance efficiency is defined as:
\textit{The degree to which a system or component performs its designated functions with minimum consumption of resources.}

Efficiency is analyzed across three key aspects: Approach Deployment Efficiency, Upgrade Deployment Efficiency, and Execution Efficiency. Unlike other themes, efficiency is not classified based on smart contract components, as it depends heavily on the gas consumption of the entire upgrade mechanism rather than individual components. This broader approach allows for a more comprehensive assessment of how various upgrade strategies affect overall resource usage and operational costs.

\subsubsection{Approach Deployment Efficiency}

Approach Deployment Efficiency refers to the initial setup cost, including the number of base contracts needed to establish each upgrade mechanism. We categorize this efficiency into four levels, where very high indicates a cost-effective setup and low represents a more resource-intensive, costly setup.

\begin{itemize}
    \item \textbf{Very High Approach Deployment Efficiency:}
    \begin{itemize}
        \item \textit{Contract Migration}: Involves minimal initial setup without proxies, making it cost-effective in deployment~\cite{hackernoon2020,li2024characterizing}.
        \item \textit{Metamorphic Contracts (CREATE2)}: Simple self-destruction and recreation require minimal setup, keeping initial deployment highly efficient~\cite{openzeppelin2020, li2024characterizing, huang2024sword}.
        \item \textit{Strategy Pattern}: Lightweight setup focusing on strategy contracts only, making deployment cost-effective while enabling selective updates\cite{medium2024proxy,cryptomarketpool2022upgrade, ebrahimi2024large}.
    \end{itemize}
    
    \item \textbf{High Approach Deployment Efficiency:}
    \begin{itemize}
        \item \textit{Data Separation Approaches (Eternal Storage and Unstructured Storage)}: Setup costs are moderate due to the need for separate storage contracts but remain efficient compared to multi-layered proxies~\cite{cryptomarketpool2022upgrade,trailofbits2018,indorse2018,securityboulevard2018}.
        \item \textit{UUPS Proxy}: Includes the upgrade mechanism within the logic contract, resulting in moderate initial setup costs to ensure state and logic are properly aligned~\cite{openzeppelin2020,logrocket2024uups, quicknode2024introduction,moralis2022,scsfg2023}.
    \end{itemize}
    
    \item \textbf{Moderate Approach Deployment Efficiency:}
    \begin{itemize}
        \item \textit{Basic Proxy and Delegate Proxy Approaches (Basic Proxy, EIP-897 Delegate Proxy, EIP-1967 Standard Storage Slots)}: The requirement to deploy both proxy and logic contracts incurs moderate costs but ensures a solid foundation for future upgrades~\cite{quicknode2024introduction,orucc2022context,klinger2020upgradeability,lopez2022upgradeable}.
    \end{itemize}
    
    \item \textbf{Low Approach Deployment Efficiency:}
    \begin{itemize}
        \item \textit{Hybrid Approaches (Proxy with Eternal Storage, Proxy Inherited Storage, Proxy Unstructured Storage)}: Dual-proxy configurations and separate storage management add initial complexity and higher gas costs~\cite{orucc2022context,mvpworkshop2021,flolio2022}.
        \item \textit{Beacon Proxy}: Requires multiple contracts, including a beacon and proxy, making it one of the most resource-intensive to set up initially~\cite{wasnik2024proxy,ebrahimiupc}.
        \item \textit{Diamond Pattern (EIP-2535)}: The multi-facet setup for separate storage management requires extensive initial gas costs, especially when deploying a large number of facets~\cite{eip2535diamonds2022,soliditydeveloper2024}.
        \item \textit{EIP-1538 (Transparent Contract Standard)}: Complex function mappings and rigorous admin verifications increase setup costs, making it among the least efficient for initial deployment~\cite{eip1538,banerjee2022}.
    \end{itemize}
\end{itemize}

\subsubsection{Upgrade Deployment Efficiency}

Upgrade Deployment Efficiency measures the gas costs incurred when deploying a new contract version during upgrades. Similar to Approach Deployment Efficiency, we categorize this into four levels, where very high represents cost-effective upgrades, and low indicates high gas costs and resource use.
\begin{itemize}
    \item \textbf{Very High Upgrade Deployment Efficiency:}
    \begin{itemize}
        \item \textit{Strategy Pattern}: Deploying only specific strategy contracts results in minimal gas usage for upgrades. This approach allows efficient redeployment without modifying core contracts by isolating the logic within individual strategies~\cite{medium2024proxy, cryptomarketpool2022upgrade, ebrahimi2024large}.
    \end{itemize}
    
    \item \textbf{High Upgrade Deployment Efficiency:}
    \begin{itemize}
        \item \textit{Basic Proxy, EIP-897 Delegate Proxy, EIP-1967 Standard Storage Slots}: Only the logic contract needs to be redeployed, and updating the implementation address in the proxy involves minimal costs. Standardized storage slots (EIP-1967) and \texttt{delegatecall} used in these proxies add simplicity, avoiding extensive authorization or setup~\cite{quicknode2024introduction,klinger2020upgradeability,amri2023review}.
        \item \textit{Data Separation Approaches (Eternal Storage and Unstructured Storage)}: Deployment costs are low, as these approaches only redeploy the logic contract and rely on separated storage~\cite{cryptomarketpool2022upgrade,trailofbits2018,indorse2018}.
    \end{itemize}
    
    \item \textbf{Moderate Upgrade Deployment Efficiency:}
    \begin{itemize}
        \item \textit{UUPS Proxy}: The upgrade functionality resides within the logic contract, requiring a complete redeployment of the logic contract each time, increasing gas consumption compared to basic proxies.
        \item \textit{Transparent Proxy}: Additional gas costs are incurred due to admin checks and the need to distinguish between admin and user roles, adding to the complexity of upgrades~\cite{logrocket2024uups,quicknode2024introduction,moralis2022}.
        \item \textit{Diamond Pattern (EIP-2535)}: While facets allow selective redeployment, managing multiple facets and ensuring consistent storage layout introduces moderate upgrade costs~\cite{eip2535diamonds2022,soliditydeveloper2024}.
        \item \textit{EIP-1538 (Transparent Contract Standard)}: This approach requires extensive function mappings and admin verifications, resulting in higher upgrade costs~\cite{eip1538,banerjee2022}.
        \item \textit{Hybrid Approaches (Proxy with Eternal Storage, Proxy Inherited Storage, Proxy Unstructured Storage)}: Dual-proxy configurations add some complexity but remain manageable due to separated data, keeping costs moderate~\cite{orucc2022context,mvpworkshop2021,flolio2022}.
    \end{itemize}
    
    \item \textbf{Low Upgrade Deployment Efficiency:}
    \begin{itemize}
        \item \textit{Contract Migration and Metamorphic Contracts (CREATE2)}: Redeploying a new version involves significant resources due to manual state migration requirements. Contract migration necessitates data migration and user notification, while Metamorphic Contracts require self-destruction and recreation, leading to additional state management costs~\cite{openzeppelin2020,salehi2022not,huang2024sword,hackernoon2020,li2024characterizing}.
    \end{itemize}
\end{itemize}

\subsubsection{Execution Efficiency}

Execution Efficiency evaluates the transaction cost of interacting with or executing the logic contract after an upgrade. Approaches using \texttt{delegatecall} typically result in lower gas costs compared to \texttt{call}-based methods. We classify execution efficiency into four levels, where very high indicates highly efficient interactions with minimal transaction costs, and low reflects higher gas costs and less efficient execution.

\begin{itemize}
    \item \textbf{Very High Execution Efficiency:}
    \begin{itemize}
        \item \textit{Contract Migration}: As this approach interacts directly without \texttt{delegatecall} or additional proxy layers, it minimizes gas usage during execution, making it highly efficient~\cite{hackernoon2020,collins2021upgrading,li2024characterizing}.
        \item \textit{Metamorphic Contracts}: These contracts benefit from direct interactions post-deployment without extra execution layers, achieving very low gas usage~\cite{openzeppelin2020,salehi2022not,huang2024sword}.
    \end{itemize}
    
    \item \textbf{High Execution Efficiency:}
    \begin{itemize}
        \item \textit{Basic Proxy and Delegate Proxy Approaches (Basic Proxy, UUPS Proxy, EIP-897 Delegate Proxy, EIP-1967 Standard Storage Slots)}: Execution relies on \texttt{delegatecall}, which maintains high efficiency for contract interactions while preserving state~\cite{klinger2020upgradeability,li2024characterizing,lopez2022upgradeable,amri2023review}.
        \item \textit{Strategy Pattern}: Direct strategy calls offer high efficiency in execution; however, accessing past state through external calls can slightly increase gas costs but remain within acceptable levels~\cite{medium2024proxy,cryptomarketpool2022upgrade,ebrahimi2024large}.
    \end{itemize}
    
    \item \textbf{Moderate Execution Efficiency:}
    \begin{itemize}
        \item \textit{Data Separation Approaches (Eternal Storage and Unstructured Storage)}: Execution relies on \texttt{call}-based access to storage contracts, adding moderate gas usage due to indirect data access and increased interactions~\cite{trailofbits2018,indorse2018,salehi2022not,stackexchange2019,securityboulevard2018}.
        \item \textit{Beacon Proxy}: Gas costs increase due to additional interactions with the beacon during execution, especially when coordinating multiple proxies~\cite{klinger2020upgradeability,li2024characterizing,lopez2022upgradeable,amri2023review}.
        \item \textit{Transparent Proxy}: Selector checks to distinguish between admin and user calls add notable gas usage, keeping it less efficient than simpler proxy options~\cite{logrocket2024uups,quicknode2024introduction,moralis2022}.
    \end{itemize}
    
    \item \textbf{Low Execution Efficiency:}
    \begin{itemize}
        \item \textit{Diamond Pattern (EIP-2535)}: Execution involves delegating to multiple facets via a dispatcher, adding significant gas overhead, especially in systems with complex storage needs~\cite{eip2535diamonds2022,soliditydeveloper2024}.
        \item \textit{Hybrid Approaches (Proxy with Eternal Storage, Proxy Inherited Storage, Proxy Unstructured Storage)}: Dual-proxy structures increase gas usage for execution due to storage separation and added complexity~\cite{orucc2022context,mvpworkshop2021,flolio2022}.
        \item \textit{EIP-1538 (Transparent Contract Standard)}: Each call requires admin verification, resulting in high storage access costs and low efficiency for frequent interactions~\cite{eip1538,banerjee2022}.
    \end{itemize}
\end{itemize}

\subsubsection{Summary of Efficiency Rankings}

\begin{figure}[t]
    \centering
    \includegraphics[width=\textwidth]{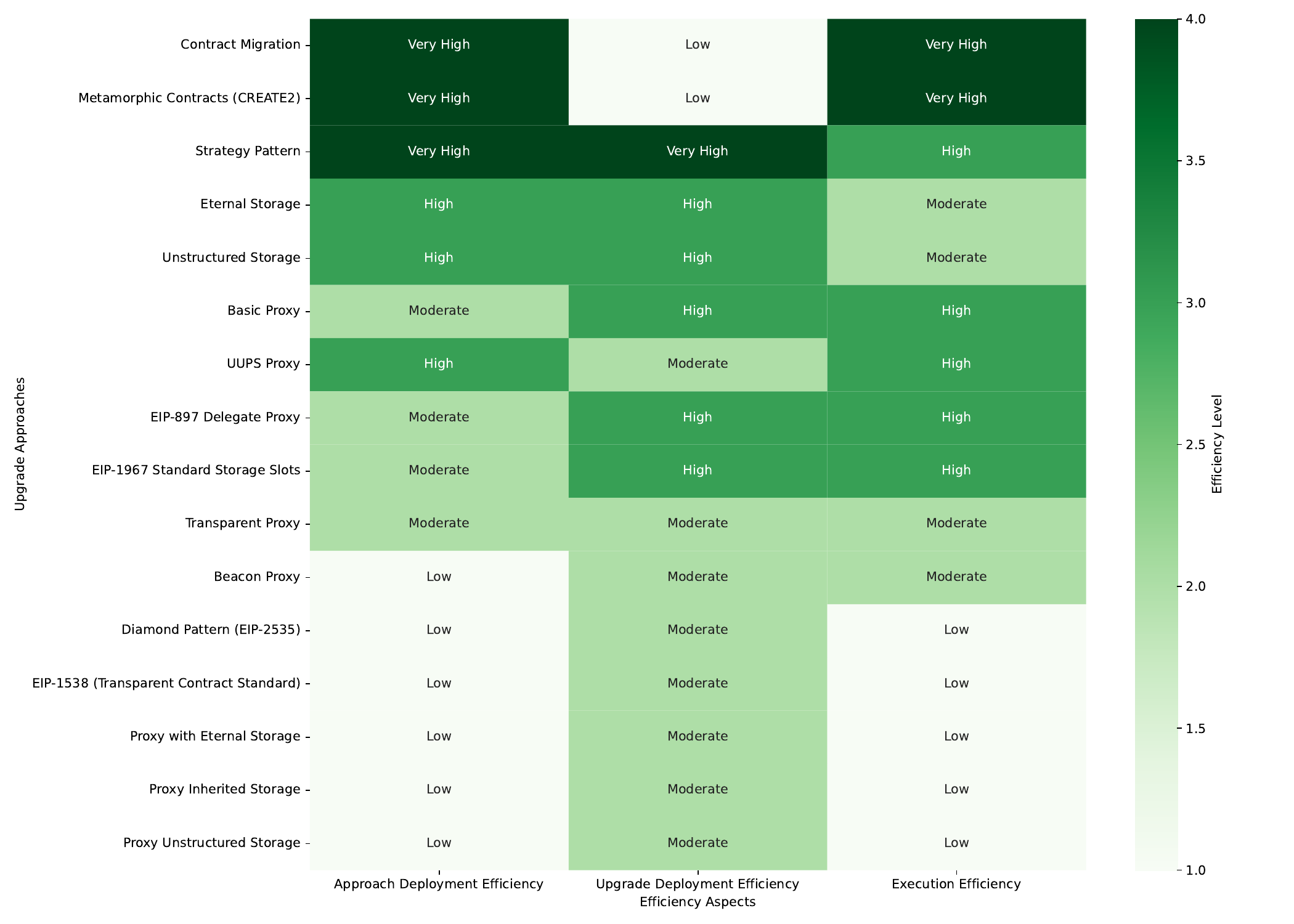} 
    \caption{Efficiency levels for smart contract upgrade approaches}
    \label{fig:RQ3-eff}
\end{figure}

As illustrated in Figure~\ref{fig:RQ3-eff}, the heatmap integrates efficiency levels across a spectrum of smart contract upgrade mechanisms. The X-axis represents the three efficiency dimensions: Upgrade Deployment Efficiency, Execution Efficiency, and Approach Deployment Efficiency, while the Y-axis lists each smart contract upgrade approach. A color gradient from light to dark tones indicates efficiency levels, where darker shades represent higher efficiency and lighter shades signify lower efficiency.
Using the average efficiency score for each approach, we categorized their overall efficiency as follows:

\begin{itemize}
    \item \textbf{Low Efficiency:}
    \begin{itemize}
        \item Approaches like \textit{Diamond Pattern (EIP-2535)}, \textit{EIP-1538 (Transparent Contract Standard)}, \textit{Proxy with Eternal Storage}, \textit{Proxy Inherited Storage}, and \textit{Proxy Unstructured Storage} exhibit low efficiency. These methods introduce complex configurations and high gas costs, especially during upgrades and initial deployments.
    \end{itemize}
    
    \item \textbf{Moderate Efficiency:}
    \begin{itemize}
        \item \textit{Beacon Proxy} and \textit{Transparent Proxy} are classified under moderate efficiency. While these approaches provide some flexibility in upgrade deployment, they incur additional gas costs during execution due to checks for admin and user distinctions or beacon interactions.
    \end{itemize}
    
    \item \textbf{High Efficiency:}
    \begin{itemize}
        \item Approaches including \textit{Contract Migration}, \textit{Metamorphic Contracts (CREATE2)}, \textit{Basic Proxy}, \textit{UUPS Proxy}, \textit{EIP-897 Delegate Proxy}, \textit{EIP-1967 Standard Storage Slots}, \textit{Eternal Storage}, and \textit{Unstructured Storage} are classified as highly efficient. These methods balance deployment and execution costs with straightforward upgrade paths and efficient interaction mechanisms.
    \end{itemize}
    
    \item \textbf{Very High Efficiency:}
    \begin{itemize}
        \item The \textit{Strategy Pattern} achieves very high efficiency by deploying only specific strategy contracts for updates, minimizing both deployment and execution costs. Its lightweight setup and direct strategy calls reduce gas consumption, making it an optimal choice for systems needing isolated, targeted updates without redeploying core components.
    \end{itemize}
\end{itemize}

\subsection{Security}

Security is a fundamental aspect of smart contract upgradeability, ensuring that only authorized changes occur and the system remains resilient against attacks. According to ISO/IEC 25010, security is defined as:
\textit{The degree to which a product or system protects information and data so that persons or other products or systems have the degree of data access appropriate to their types and levels of authorization.}

We examine security vulnerabilities in smart contract upgradeability across three categories: Logic Vulnerabilities, Storage Vulnerabilities, and Execution Flow Vulnerabilities.

\subsubsection{Logic Vulnerabilities}

Logic vulnerabilities are critical risks that arise from handling contract logic upgrades or delegate operations.

\begin{itemize}
    \item \textbf{Delegatecall Misuse}
    \begin{itemize}
        \item \textit{Affected Approaches}: Proxy-Based Approaches, including Basic Proxy, UUPS Proxy, Beacon Proxy, and EIP-897, as well as \textit{Diamond Pattern (EIP-2535)} due to \texttt{delegatecall} reliance in facets~\cite{openzeppelin2020,trailofbits2018,orucc2022context,lopez2022upgradeable}.
        \item \textit{Partial Mitigation}: The Transparent Proxy Pattern reduces this risk by segregating administrative functions from user functions. However, if the admin account is compromised, this mitigation is less effective~\cite{openzeppelin2020,certik2022,scsfg2023}.
    \end{itemize}
    
    \item \textbf{Function Selector Clashes}
    \begin{itemize}
        \item \textit{Affected Approaches}: Proxy-Based Approaches, especially Basic Proxy, Beacon Proxy, and EIP-897, are susceptible to selector clashes. \textit{Diamond Pattern} increases clash risks due to dynamic function handling~\cite{jeiwan2021,certik2022,wasnik2024proxy}.
        \item \textit{Partial Mitigation}: UUPS Proxy reduces the risk by embedding the upgrade function within the logic contract~\cite{bodell2023proxy,eip1822}. The Transparent Proxy also minimizes this risk by separating administrative functions from user functions~\cite{ebrahimiupc,quicknode2024introduction}.
    \end{itemize}
    
    \item \textbf{Upgrade Initialization Risks and Bricking}
    \begin{itemize}
        \item \textit{Affected Approaches}: UUPS Proxy is particularly vulnerable, as it requires upgrade functionality within the logic contract~\cite{li2024characterizing,openzeppelin2020,logrocket2024uups,qasse2024immutable}. Metamorphic Contracts (CREATE2) also face bricking risks if redeployed incorrectly~\cite{li2024characterizing, huang2024sword}. The Diamond Pattern (EIP-2535) is similarly affected, as its upgrade function resides in the logic rather than in the proxy, increasing the risk of improper upgrades~\cite{bodell2023proxy,eip2535diamonds2022}.
    \end{itemize}
\end{itemize}

\subsubsection{Storage Vulnerabilities}

Storage vulnerabilities, such as storage collisions, layout changes across versions, and unauthorized access, affect data integrity.

\begin{itemize}
    \item \textbf{Storage Collisions}
    \begin{itemize}
        \item \textit{Affected Approaches}: Proxy-Based Approaches like Basic Proxy, Beacon Proxy, and EIP-897, as well as \textit{Diamond Pattern} due to shared storage among facets~\cite{moralis2022,openzeppelin2020,jeiwan2021,trailofbits2018,ethereumdeveloper2021upgrade,quicknode2024introduction}.
        \item \textit{Partial Mitigation}: Unstructured Storage or Data Separation patterns mitigate this risk by separating storage from the logic contract~\cite{bodell2023proxy,klinger2020upgradeability,orucc2022context}.
    \end{itemize}
    
    \item \textbf{Storage Layout Changes Between Versions}
    \begin{itemize}
        \item \textit{Affected Approaches}: All Proxy-Based Approaches face this risk.
        \item \textit{Mitigation}: Eternal Storage or Unstructured Storage Patterns in Data Separation effectively mitigate this risk by keeping storage in a separate contract.
    \end{itemize}
    
    \item \textbf{Unauthorized Access to Storage}
    \begin{itemize}
        \item \textit{Affected Approaches}: Data Separation approaches without strict access control, Hybrid Approaches without adequate restrictions~\cite{ethereum2024upgrading, trailofbits2018,indorse2018,securityboulevard2018}.
        \item \textit{Mitigation}: Implementing strict access controls and using role-based access control (RBAC) frameworks.
    \end{itemize}
\end{itemize}

\subsubsection{Execution Flow Vulnerabilities}

Execution flow vulnerabilities relate to risks in function call control and upgrade mechanisms.

\begin{itemize}
    \item \textbf{Unauthorized Upgrades}
    \begin{itemize}
        \item \textit{Affected Approaches}: Beacon Proxy, Proxy Patterns with centralized admin control, UUPS Proxy, and Metamorphic Contracts (CREATE2)~\cite{ethereum2024upgrading, trailofbits2018,indorse2018,securityboulevard2018,hackernoon2020,lopez2022upgradeable,Bloo2018,blockchains2024,certik2022,scsfg2023}.
    \end{itemize}
    
    \item \textbf{Centralization Risks}
    \begin{itemize}
        \item \textit{Affected Approaches}: Beacon Proxy, Proxy with Centralized Admin~\cite{openzeppelin2020,certik2022,Bloo2018,runtime2022,ebrahimi2024large}.
    \end{itemize}
\end{itemize}

\subsubsection{Summary of Security Implications by Aspect}

Each approach's security risks depend on the specific vulnerabilities addressed in Logic, Storage, and Execution Flow, as shown in Table~\ref{tab:RQ3-S}. Below is a ranking for each category based on their security implications:

\begin{itemize}
    \item \textbf{Logic Vulnerabilities Ranking}
    \begin{itemize}
        \item \textit{Low Vulnerability}: Contract Migration, Strategy Pattern
        \item \textit{Moderate Vulnerability}: Transparent Proxy, EIP-1967, Data Separation
        \item \textit{High Vulnerability}: Basic Proxy, UUPS Proxy, Diamond Pattern
    \end{itemize}
    
    \item \textbf{Storage Vulnerabilities Ranking}
    \begin{itemize}
        \item \textit{Low Vulnerability}: Unstructured Storage, Data Separation
        \item \textit{Moderate Vulnerability}: Transparent Proxy, EIP-1967, Strategy Pattern
        \item \textit{High Vulnerability}: Basic Proxy, Contract Migration, Diamond Pattern
    \end{itemize}
    
    \item \textbf{Execution Flow Vulnerabilities Ranking}
    \begin{itemize}
        \item \textit{Low Vulnerability}: Contract Migration, Strategy Pattern
        \item \textit{Moderate Vulnerability}: Transparent Proxy, EIP-1967
        \item \textit{High Vulnerability}: Diamond Pattern, Beacon Proxy, Metamorphic Contracts (CREATE2)
    \end{itemize}
\end{itemize}

\textbf{Note:} The security of each approach depends heavily on proper implementation, rigorous testing, and adherence to best practices. Even approaches with higher inherent risks can be made secure through diligent development and comprehensive security measures.

\begin{table}[]
\centering
\caption{Summary of Security Vulnerabilities in Smart Contract Upgradeability Approaches}
\label{tab:RQ3-S}
\resizebox{\textwidth}{!}{%
\begin{tabular}{ccccc}
\textbf{Security Category} &
  \textbf{Vulnerability} &
  \textbf{Affected Approaches} &
  \textbf{Partial Mitigations} &
  \textbf{Severity Ranking} \\
\multirow{3}{*}{\textbf{Logic Vulnerabilities}} &
  Delegatecall Misuse &
  \begin{tabular}[c]{@{}c@{}}Proxy-Based Approaches, \\ Diamond Pattern\end{tabular} &
  Transparent Proxy &
  \begin{tabular}[c]{@{}c@{}}High \\ (if delegatecall mishandled)\end{tabular} \\
 &
  Function Selector Clashes &
  \begin{tabular}[c]{@{}c@{}}Basic Proxy, Beacon Proxy, EIP-897, \\ Diamond Pattern, EIP-1538\end{tabular} &
  \begin{tabular}[c]{@{}c@{}}Transparent Proxy, \\ UUPS Proxy\end{tabular} &
  Moderate to High \\
 &
  \begin{tabular}[c]{@{}c@{}}Upgrade Initialization \\ Risks and Bricking\end{tabular} &
  \begin{tabular}[c]{@{}c@{}}UUPS Proxy, \\ Metamorphic Contracts\end{tabular} &
  \begin{tabular}[c]{@{}c@{}}Transparent Proxy, \\ EIP-1967\end{tabular} &
  Moderate to High \\
\multirow{3}{*}{\textbf{Storage Vulnerabilities}} &
  Storage Collisions &
  \begin{tabular}[c]{@{}c@{}}Basic Proxy, Beacon Proxy, \\ EIP-897, Diamond Pattern, \\ Inherited Storage\end{tabular} &
  \begin{tabular}[c]{@{}c@{}}Unstructured Storage, \\ Data Separation, \\ EIP-1967\end{tabular} &
  Moderate to High \\
 &
  Layout Changes &
  Proxy-Based Approaches &
  Hybrid Approaches &
  Moderate \\
 &
  \begin{tabular}[c]{@{}c@{}}Unauthorized \\ Storage Access\end{tabular} &
  \begin{tabular}[c]{@{}c@{}}Data Separation types, \\ Hybrid Approaches, \\ Metamorphic Contracts\end{tabular} &
  - &
  High if poorly managed \\
\multirow{2}{*}{\textbf{Execution Flow}} &
  \begin{tabular}[c]{@{}c@{}}Unauthorized \\ Upgrades\end{tabular} &
  \begin{tabular}[c]{@{}c@{}}Beacon Proxy, Transparent Proxy, \\ UUPS, Metamorphic Contracts\end{tabular} &
  - &
  Moderate to High \\
 &
  \begin{tabular}[c]{@{}c@{}}Centralization \\ Risks\end{tabular} &
  \begin{tabular}[c]{@{}c@{}}Beacon Proxy, Proxy with \\ Centralized Admin\end{tabular} &
  - &
  Moderate
\end{tabular}%
}
\end{table}
\subsection{Usability}

Usability is essential in designing smart contract upgrade mechanisms, as it directly impacts user experience and trust in the system. According to ISO/IEC 25010, usability is defined as:

\textit{The degree to which specified users can use a system to achieve specified goals with effectiveness, efficiency, and satisfaction in a specified context of use.}

In smart contracts, usability is evaluated based on two key aspects: Interaction Simplicity and Transparency. These aspects ensure that users can easily interact with the system and clearly understand its operations.

\subsubsection{Interaction Simplicity of Smart Contract Upgrade Approaches}

Interaction Simplicity refers to how easily users can continue to interact with the contract after an upgrade, particularly regarding whether the contract address stays consistent or needs updating. We rank this aspect into three levels, where low indicates significant challenges for users, and high represents seamless interactions and minimal disruptions.

\begin{itemize}
    \item \textbf{Low Interaction Simplicity:}
    \begin{itemize}
        \item \textit{Contract Migration}: Each upgrade involves a new contract address, requiring users to switch addresses manually and migrate their data or assets. This complexity in maintaining interaction with the current contract significantly reduces usability~\cite{hackernoon2020,collins2021upgrading,ebrahimiupc}.
        \item \textit{Data Separation Approaches (Inherited Storage, Eternal Storage, Unstructured Storage)}: Similar to Contract Migration, these approaches often require separate data and logic contracts. Users must interact with new addresses post-upgrade, making interaction cumbersome and less intuitive~\cite{ethereum2024upgrading,cryptomarketpool2022upgrade,indorse2018,stackexchange2019}.
    \end{itemize}
    
    \item \textbf{Moderate Interaction Simplicity:}
    \begin{itemize}
        \item \textit{Metamorphic Contracts (CREATE2)}: While the address remains stable, users may encounter downtime and potential state resets during redeployment. Though interactions are consistent, users may experience disruptions affecting interaction simplicity~\cite{openzeppelin2020,salehi2022not,ebrahimiupc}.
        \item \textit{Strategy Pattern}: The main contract remains stable, simplifying interactions. However, changes to underlying strategy contracts can modify specific functionalities, requiring slight user adaptation to potential changes in contract behavior~\cite{ebrahimiupc,li2024characterizing,openzeppelin2021security,salehi2022not}.
    \end{itemize}
    
    \item \textbf{High Interaction Simplicity:}
    \begin{itemize}
        \item \textit{Proxy Approaches}: The proxy maintains a stable address, allowing users to interact consistently without changing their methods. Users experience seamless upgrades as the interface remains unchanged~\cite{blockmagnates2022upgradability,klinger2020upgradeability,lopez2022upgradeable,huang2024sword}.
        \item \textit{Hybrid Approaches (Proxy with Eternal Storage, Proxy Inherited Storage, Proxy Unstructured Storage)}: These approaches use proxy mechanisms, preserving a constant address, which ensures continuity and high interaction simplicity~\cite{ebrahimiupc,klinger2020upgradeability, bui2021evaluating,mvpworkshop2021}.
    \end{itemize}
\end{itemize}

\subsubsection{Transparency of Smart Contract Upgrade Approaches}

Transparency refers to the extent to which users are aware of upgrades and understand their impact on contract functionality. Similar to Interaction Simplicity, we rank transparency into three levels, where low indicates limited user awareness and understanding, and high signifies full transparency with clear communication and understanding of the upgrade's effects.

\begin{itemize}
    \item \textbf{Low Transparency:}
    \begin{itemize}
        \item \textit{Contract Migration}: Users must be informed of address changes manually, with a high likelihood of continuing interactions with outdated contracts if not fully notified~\cite{hackernoon2020,collins2021upgrading,ebrahimiupc}.
        \item \textit{Data Separation Approaches (Inherited Storage, Eternal Storage, Unstructured Storage)}: These approaches lack inherent notification mechanisms. Without a registry to point to the latest contract, users are often unaware of upgrades~\cite{indorse2018,salehi2022not}.
    \end{itemize}
    
    \item \textbf{Moderate Transparency:}
    \begin{itemize}
        \item \textit{Proxy Approaches (Basic Proxy, UUPS, EIP-897 Delegate Proxy, Transparent Proxy)}: Users can detect upgrades by reviewing proxy transactions, but these require active checking. Users may not be directly informed of updates without prior knowledge of where to look~\cite{blockmagnates2022upgradability,certik2022,amri2023review,huang2024sword,Bloo2018}.
        \item \textit{Hybrid Approaches (Proxy with Eternal Storage, Proxy Inherited Storage, Proxy Unstructured Storage)}: Similar to basic proxies, users can determine upgrades through transaction records, though direct awareness depends on proactive checking~\cite{ebrahimiupc,klinger2020upgradeability,bui2021evaluating}.
    \end{itemize}
    
    \item \textbf{High Transparency:}
    \begin{itemize}
        \item \textit{EIP-1967 Standard Storage Slots, Diamond Pattern (EIP-2535), EIP-1538 (Transparent Contract Standard), Beacon Proxy}: These approaches emit events during upgrades, notifying users of changes and the new implementation address. This proactive communication enhances user awareness of modifications and preserves trust through transparent upgrade mechanisms~\cite{eip1967,eip1538,eip2535diamonds2022,soliditydeveloper2024,qasse2024immutable}.
    \end{itemize}
\end{itemize}

\subsubsection{Summary of Usability Rankings}

\begin{figure}[t]
    \centering
    \includegraphics[width=\textwidth]{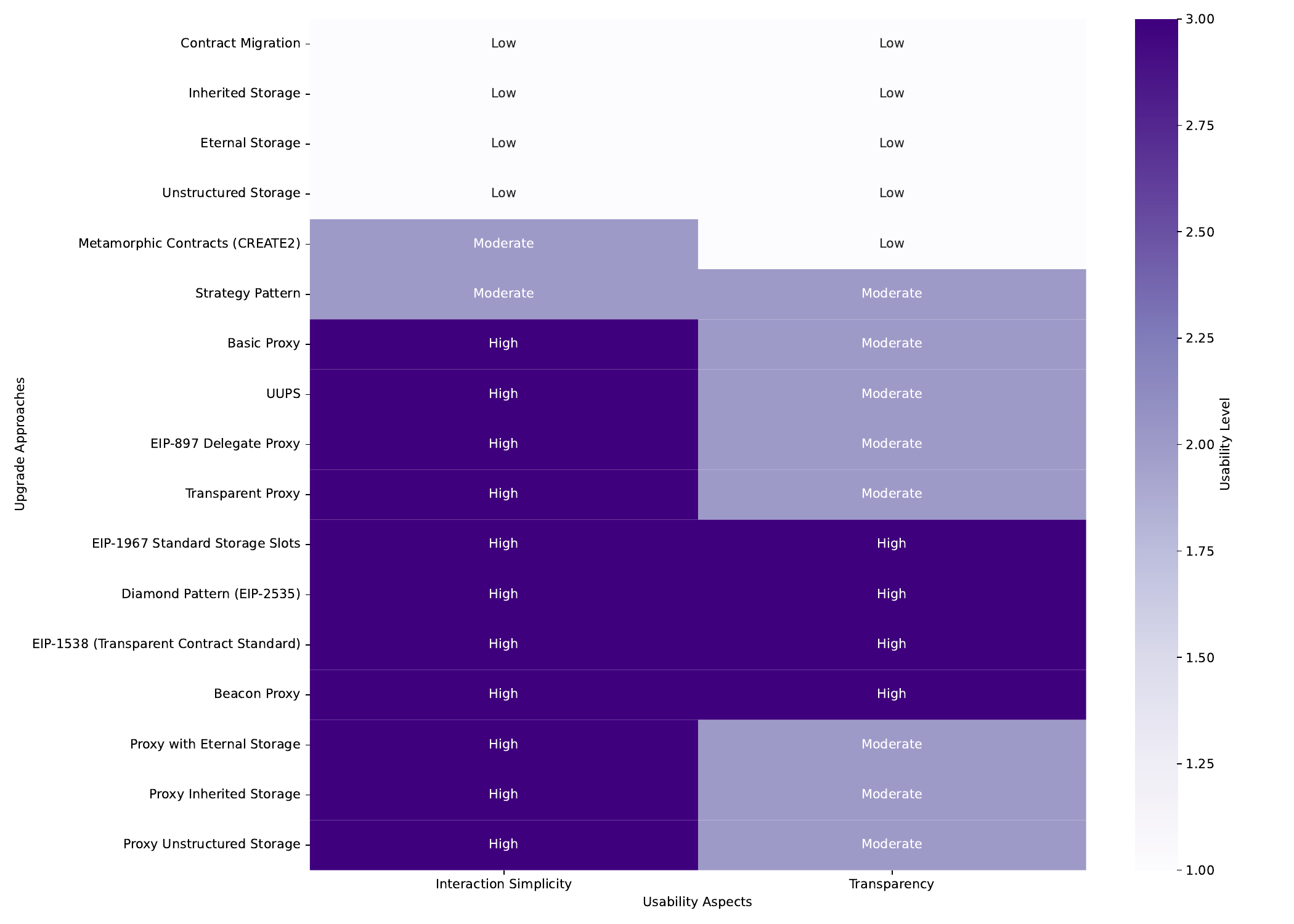} 
    \caption{Usability levels for smart contract upgrade approaches}
    \label{fig:RQ3-use}
\end{figure}

Figure~\ref{fig:RQ3-use} illustrates that the heatmap integrates usability levels across a spectrum of smart contract upgrade mechanisms. The X-axis represents the two primary usability dimensions—Interaction Simplicity and Transparency—while the Y-axis lists each smart contract upgrade approach. A color gradient from light to dark tones indicates usability levels, where darker shades represent higher usability and lighter shades signify lower usability.
Using the average usability score for each approach, we categorized their overall usability as follows:

\begin{itemize}
    \item \textbf{Low Usability:}
    \begin{itemize}
        \item Approaches like \textit{Contract Migration}, \textit{Inherited Storage}, \textit{Eternal Storage}, and \textit{Unstructured Storage} exhibit low usability. These methods introduce interaction complexity as they require users to interact with different addresses post-upgrade and lack clear mechanisms for notifying users of upgrades.
    \end{itemize}
    
    \item \textbf{Moderate Usability:}
    \begin{itemize}
        \item The \textit{Metamorphic Contracts (CREATE2)} and \textit{Strategy Pattern} approaches fall under moderate usability. While Metamorphic Contracts maintain a stable address, they may involve downtime and potential state resets during redeployment. The Strategy Pattern retains a stable main contract address, simplifying interaction, though underlying strategy changes may require slight user adaptation.
    \end{itemize}
    
    \item \textbf{High Usability:}
    \begin{itemize}
        \item A variety of approaches, including \textit{Basic Proxy}, \textit{UUPS}, \textit{EIP-897 Delegate Proxy}, \textit{Transparent Proxy}, \textit{EIP-1967 Standard Storage Slots}, \textit{Diamond Pattern (EIP-2535)}, \textit{EIP-1538 (Transparent Contract Standard)}, \textit{Beacon Proxy}, \textit{Proxy with Eternal Storage}, \textit{Proxy Inherited Storage}, and \textit{Proxy Unstructured Storage} are classified as highly usable. These approaches offer stable addresses, enabling users to interact seamlessly post-upgrade and benefit from moderate to high transparency.
    \end{itemize}
\end{itemize}

\subsection{RQ3 Answer}
\begin{figure}[t]
    \centering
    \includegraphics[width=\textwidth]{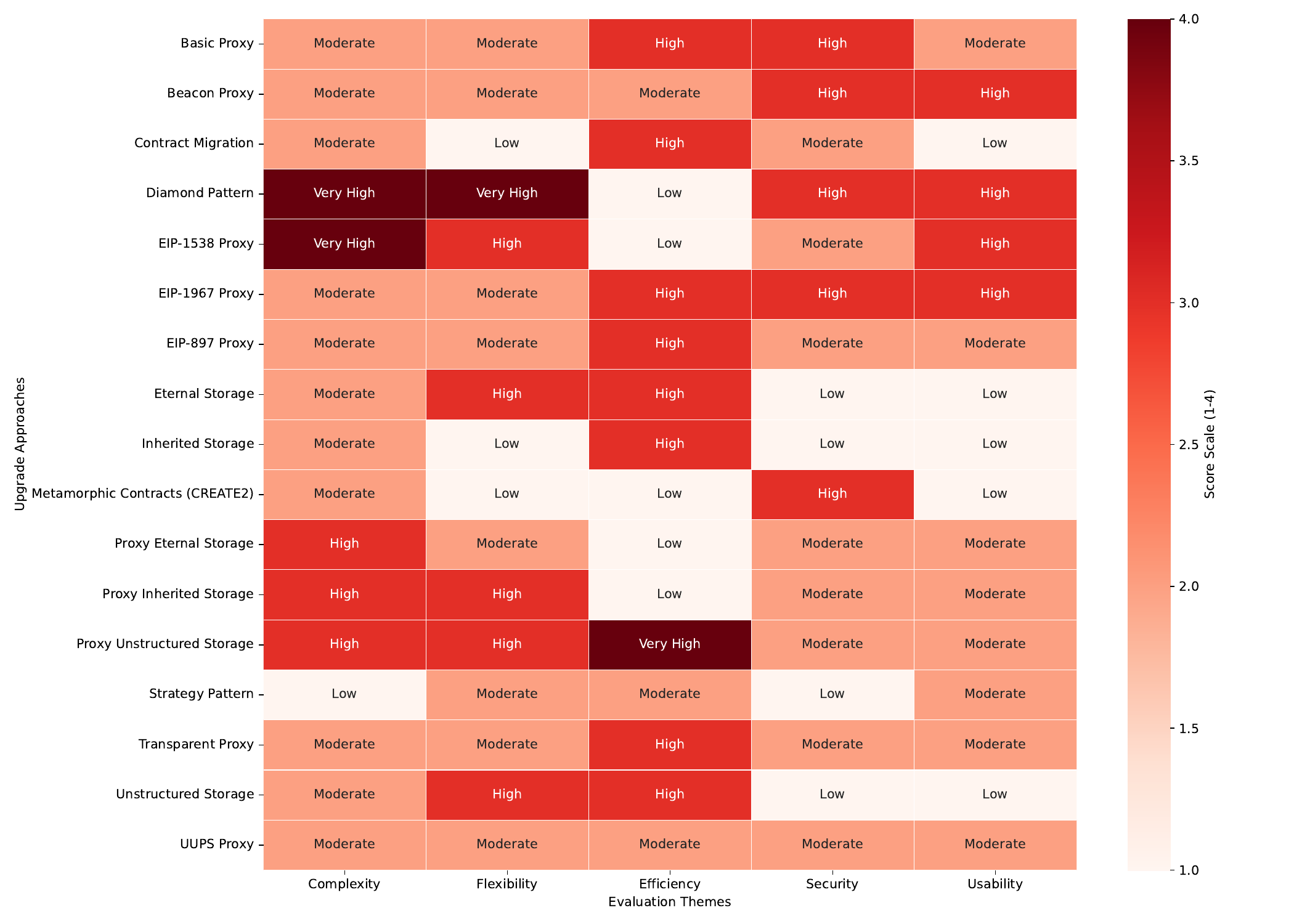} 
    \caption{Summary of smart contract upgrade approaches accross all themes}
    \label{fig:RQ3-sum}
\end{figure}

To address RQ3: What are the benefits and limitations of each upgrading approach?, we evaluated each approach based on attributes such as complexity, flexibility, efficiency, security, and usability. Figure~\ref{fig:RQ3-sum} presents a heatmap that offers a comprehensive overview of the performance of different smart contract upgrade approaches across these themes. Choosing the right approach often means balancing these themes to fit a project's unique needs. For projects that prioritize simplicity and safety, Strategy Pattern and Unstructured Storage are strong options. The Strategy Pattern stands out for its low complexity and minimal security risks, making it ideal for applications where straightforward updates are needed. Unstructured Storage, though more complex, offers high flexibility and good efficiency while keeping security risks low. However, it has lower usability due to its complex implementation.
For projects looking for a balance across most themes, UUPS Proxy and Transparent Proxy offer reliable solutions. UUPS Proxy combines low complexity with high efficiency, providing moderate flexibility and usability, though it needs careful handling to address some security risks. Transparent Proxy has moderate complexity and security risks but delivers strong efficiency and usability, making it a practical and widely used choice. Both proxies keep the contract address the same during upgrades, so users do not need to update their interactions, which adds to usability.

For projects needing high flexibility, especially with large and complex contracts, Diamond Pattern excels by allowing modular upgrades. However, this comes with higher complexity and security risks, requiring skilled developers and thorough security practices. Basic Proxy and EIP-1967 Proxy offer a balanced mix of moderate complexity, flexibility, and security, along with good efficiency and usability, making them suitable for various types of projects. Ultimately, selecting the right upgrade approach involves understanding the trade-offs between themes. The heatmap in Figure~\ref{fig:RQ3-sum} serves as a helpful visual guide to compare these approaches and find the best fit based on the specific needs of the project.

\section{Discussion}
\label{sec:discussion}
This section synthesizes insights on smart contract upgradeability, focusing on governance models, lifecycle management, and the decision "to upgrade or not." While upgrades offer adaptability, they do not inherently guarantee improved security or usability. Each theme is supported by case studies and best practices, highlighting the complexities and trade-offs of smart contract upgradeability.

\subsection{Governance Models in Upgradeability}
Smart contract governance utilizes on-chain rules to establish decision-making processes and manage upgrades with the goals of adaptability, security, and transparency. Governance structures significantly influence user trust, as they determine who controls the upgrade process and how transparently changes are implemented. However, governance in smart contracts faces critical challenges, particularly in balancing decentralization with effective oversight and quick response to urgent issues.

\subsubsection{Centralized Governance Models}
Centralized models, commonly used in patterns like the Transparent Proxy and Beacon Proxy, consolidate upgrade control within a single administrator or a small group, allowing for rapid decision-making. For instance, OpenZeppelin centralizes control under an admin account for efficient upgrades.\footnote{\url{https://docs.openzeppelin.com/contracts/4.x/api/proxy}} While efficient, centralized governance can create vulnerabilities, as shown by the PAID Network hack. In this incident, attackers compromised the admin's private key, resulting in unauthorized token minting and significant user losses. This underscores the inherent risks of centralization, as a single point of failure can lead to catastrophic security breaches and erode user trust.\footnote{\url{https://blog.paidnetwork.com/}}

\subsubsection{Decentralized Governance Models}
Decentralized models often rely on community voting or multi-signature controls, where multiple stakeholders hold decision-making power to reduce reliance on any one entity. Compound and MakerDAO, two decentralized finance (DeFi) protocols, utilize token-based voting to empower their communities in governance decisions.\footnote{\url{https://compound.finance/}}\footnote{\url{https://makerdao.com/en/governance}} However, decentralized governance faces challenges like low voter participation and delayed responses. For instance, a critical bug in Compound once mistakenly distributed \$90 million in COMP tokens due to the delay needed to reach consensus on corrective action, revealing the limitations of decentralized models in urgent situations.\footnote{\url{https://www.icba.org/newsroom/blogs/main-street-matters/2021/11/12/the-challenges-and-risks-of-smart-contracts}} To mitigate this, Compound employs timelocks to give users time to review and object to upgrades, although this setup sacrifices responsiveness.

\subsubsection{Reliance on Oracles and External Data}
Some smart contract systems depend on oracles for off-chain data, such as asset prices in DeFi protocols. Oracles introduce vulnerabilities as they rely on external data sources that can be manipulated, compromising the contract's security and decentralization. For example, inaccurate or malicious data from oracles has led to significant losses in DeFi protocols, emphasizing the need for robust governance and verification mechanisms to manage these dependencies securely.\footnote{\url{https://corpgov.law.harvard.edu/2020/04/25/an-introduction-to-smart-contracts-and-their-potential-and-inherent-limitations}}

\subsubsection{Distinct Governance Models in Platforms}
Blockchain platforms adopt a variety of governance models to manage upgrades, balancing flexibility, decentralization, and transparency. For instance, Tezos\footnote{\url{https://tezos.com/}} enables upgrades through community voting without replacing the entire contract, promoting flexibility while maintaining decentralization. Similarly, Polkadot\footnote{\url{https://polkadot.network/}} employs an on-chain governance system where token holders and a council collaborate to propose and approve upgrades. In contrast, EOS\footnote{\url{https://eosnetwork.com/}} centralizes control by allowing select account groups to manage upgrades, raising transparency and decentralization concerns. These examples illustrate the diversity in governance structures, each with trade-offs in security, efficiency, and user trust.\footnote{\url{https://www.dutchblockchaincoalition.org/}}

\subsection{Best Practices in Governance}
Implementing governance in smart contracts requires best practices to balance security, transparency, and decentralized control:
\begin{itemize}
    \item \textbf{Define Clear Rules and Stakeholder Responsibilities:} Establishing transparent, well-documented processes for decision-making is crucial. Clearly defining roles, voting protocols, and decision criteria helps ensure a fair governance process, enhancing user trust and minimizing ambiguity.\footnote{\url{https://www.doubloin.com/what-is-smart-contract-governance-make-on-chain-decisions}}
    \item \textbf{Multi-Signature Wallets and Timelocks:} Multi-signature wallets prevent unilateral control by requiring multiple approvals for sensitive actions, like upgrades. Timelocks introduce delays that allow stakeholders to review and, if necessary, intervene. Protocols like Uniswap and Compound employ multisig setups and timelocks to balance rapid response capabilities with robust oversight.\footnote{\url{https://blog.openzeppelin.com/smart-contract-security-guidelines-4-strategies-for-safer-governance-systems}}
    \item \textbf{Continuous Auditing and Monitoring:} Regular audits are essential for identifying vulnerabilities in governance mechanisms, especially as protocols evolve. Monitoring transaction activities through alerts and off-chain event logs enables quick responses to unusual activities, improving security and operational resilience.\footnote{\url{https://www.webisoft.com/blog/best-practices-for-smart-contract-security}}
    \item \textbf{Stakeholder Engagement:} Voting systems, common in DeFi platforms, allow token holders to participate in governance. Although beneficial, these systems are vulnerable to manipulations like flash loans, where attackers temporarily gain control over voting. Ethereum.org emphasizes the need for safeguards in decentralized voting mechanisms to protect governance integrity.\footnote{\url{https://ethereum.org/en/developers/docs/smart-contracts/security}}
    \item \textbf{Emergency Stop and Fail-Safe Mechanisms:} An emergency stop allows authorized entities to pause critical operations during security incidents. These functions should ideally be decentralized to prevent misuse, providing a fail-safe in high-risk situations without relying on centralized control.\footnote{\url{https://ethereum.org/en/developers/docs/smart-contracts/security}}
    \item \textbf{Adapt and Improve Through Feedback:} Regularly reassessing governance based on user feedback and evolving needs helps refine the model. Continuous improvements align governance structures with community expectations and long-term project goals, sustaining trust and adaptability.\footnote{\url{https://www.doubloin.com/what-is-smart-contract-governance-make-on-chain-decisions}}
\end{itemize}

\subsection{Future Research Directions}
Despite the importance of governance in decentralized applications, empirical studies on governance models in smart contracts remain limited. Research is crucial to evaluate the practical implications of governance structures on security, user trust, and project stability. Specific areas for future investigation include:
\begin{itemize}
    \item \textbf{Comparative Studies of Centralized vs. Decentralized Models:} Evaluating the cost-benefit of different governance structures would provide insights into optimizing models for security and efficiency. Empirical studies could examine user trust across various structures, assessing the impact of centralized versus decentralized approaches on engagement and security.
    \item \textbf{Adaptive Timelock Mechanisms:} Adaptive timelocks that adjust based on the urgency or scale of the upgrade could help strike a balance between responsiveness and security. Research into such mechanisms, particularly in DeFi, would support flexible governance protocols that align with both community needs and security requirements.
    \item \textbf{Lifecycle Tracking and Real-Time Monitoring Tools:} The development of tools that offer real-time monitoring and lifecycle tracking for governance decisions could empower stakeholders by increasing transparency. These tools would allow the community to actively monitor governance activities, fostering a transparent and secure environment.
    \item \textbf{Stakeholder Engagement and Voting Efficiency:} Studies focused on improving voter participation and engagement, potentially through incentivized voting models, would provide actionable insights into strengthening decentralized governance. Given the challenges of voter apathy, research on optimal incentive structures could increase participation and ensure that community-driven governance remains effective and representative.
    \item \textbf{Best Practices for Oracle Governance:} With the increasing reliance on oracles in smart contract governance, research on best practices for oracle management is needed. Empirical studies examining oracle reliability, security, and how oracle data impacts contract outcomes would contribute to more robust governance structures that minimize external dependencies.
\end{itemize}

\subsection{Lifecycle Management: Ensuring Stability and Reducing Complexity}
Lifecycle management addresses how contracts evolve from deployment through updates, requiring strategic planning to maintain stability and user trust. Key approaches include single-time upgradeability, routine upgradeability through proxy patterns, and partial immutability with the option to "brick" contracts. Each has specific implications for usability, security, and complexity.

\subsubsection{Single-Time Upgradeability}
Single-time upgradeability, seen in methods like Contract Migration and Metamorphic Contracts (CREATE2), involves deploying a new contract each time an upgrade is necessary. This approach avoids complexities associated with \texttt{delegatecall} or layered proxy mechanisms. However, it requires users to adopt new contract addresses, potentially creating confusion and increasing the risk of interacting with outdated or malicious contracts.

One relevant example is Synthetix's\footnote{https://developer.synthetix.io/contracts/} migration process, in which contracts must be fully redeployed with each upgrade. This approach necessitates detailed communication with users, as failure to update addresses can result in interactions with obsolete contracts. Synthetix's approach underscores the challenge of maintaining user engagement and trust in migration models, where user adaptation costs remain a concern.

\subsubsection{Routine Upgradeability through Structured Patterns}
Routine upgradeability models allow frequent updates without altering the contract address, enhancing usability by maintaining a stable address interface. Although proxy patterns like UUPS (Universal Upgradeable Proxy Standard) are commonly associated with this model, other approaches, such as the Strategy Pattern or Diamond Standard (EIP-2535), also enable structured upgrades. In Audius,\footnote{\url{https://audius.co/}} a storage collision in a proxy pattern led to a governance hack, underscoring the importance of rigorous storage layout management in upgradeable contracts. Such incidents demonstrate the need for careful structuring in routine upgrades to prevent data collisions and unauthorized access.
\subsubsection{Partial Immutability}
Partial immutability, orbricking, allows developers to permanently remove upgradeability from a contract, effectively freezing its state. This approach can increase user trust by ensuring the contract remains stable and unalterable once it reaches maturity. However, bricking can inadvertently prevent necessary upgrades or security patches if misapplied.

\paragraph{Case Study: UUPSUpgradeable Vulnerabilities}
A notable instance of accidental loss of upgradeability involved OpenZeppelin's UUPSUpgradeable contracts. In 2021, a vulnerability in \texttt{\_authorizeUpgrade} allowed unauthorized upgrades on uninitialized contracts, bypassing security checks. This oversight impacted several projects and exposed assets to risk, which was resolved in OpenZeppelin Contracts v4.3.2 by adding an \texttt{onlyProxy} modifier to restrict upgrade calls directly on the implementation contract.\footnote{\url{https://forum.openzeppelin.com/t/uupsupgradeable-vulnerability-post-mortem/15680}}\footnote{\url{https://iosiro.com/blog/uups-proxy-security-review}}

\paragraph{Trade-Offs of Bricking}
While bricking can ensure stability, it carries the risk of blocking essential maintenance if used prematurely. Balancing the security benefits of immutability with the need for potential updates is essential for effective lifecycle management in smart contracts.

\subsubsection{Future Research Directions}
As smart contracts evolve, lifecycle tracking and version control tools tailored to blockchain's decentralized, often immutable nature are urgently needed.
\begin{itemize}
    \item \textbf{Lifecycle Tracking and Monitoring Tools:} These tools would offer real-time insights into contract states, including upgrade history, current versions, and any upcoming changes, providing users and developers with transparency across all contract stages. This visibility would allow users to make informed decisions about interacting with the contract, fostering trust and confidence in its long-term stability.
    \item \textbf{Version Control and Standardization Protocols:} Standardized protocols are essential to streamline upgrade management and support rollback capabilities, much like traditional software systems. These would enable developers to track upgrades, review prior versions, and maintain continuity across complex, long-term projects, improving security and usability.
    \item \textbf{Empirical Research on Partial Immutability and Lifecycle Strategies:} Empirical studies exploring the timing, effectiveness, and user perception of bricking can provide insight into how and when contracts should transition to a fixed state. This research could guide user adaptation across different lifecycle strategies, including the impact of bricking on user confidence and contract sustainability. Case studies of both successful and failed bricking implementations would offer practical examples, aiding developers in understanding the best timing and conditions for applying partial immutability.
\end{itemize}

\subsection{To Upgrade or Not: The Impact on User Trust and Security}
The decision to make a smart contract upgradeable is crucial, balancing the advantages of flexibility with the foundational blockchain principle of immutability. While upgradeability enables developers to address vulnerabilities, introduce new functionalities, and ensure compatibility with evolving standards, it also introduces risks that challenge traditional assumptions of security and trust within the blockchain ecosystem. This section explores the benefits and risks associated with upgradeable contracts, along with best practices and directions for future research.

\subsubsection{Benefits of Upgradeability}
\paragraph{Bug Fixes and Security Improvements}
The ability to address bugs and vulnerabilities post-deployment is a major advantage of upgradeable contracts. In decentralized finance (DeFi), where considerable value is often locked in contracts, vulnerabilities can lead to severe financial risk. Selective upgradeability, managed carefully, supports both security and efficiency by addressing security flaws without extensive disruptions to the contract's core functionality.\footnote{\url{https://ethereum.org/en/developers/docs/smart-contracts/upgrading/}}

\paragraph{Adaptability to New Features and Scalability}
Upgradeability facilitates the addition of new functionalities and supports scalability, particularly for long-term projects. Patterns such as the Diamond Pattern employ modular facets, which allow developers to expand or modify contract functions without disrupting the entire codebase. This modularity reduces costs and supports growth, as developers can gradually scale their projects without needing full replacements, making it ideal for applications with evolving requirements.\footnote{\url{https://blog.openzeppelin.com/the-state-of-smart-contract-upgrades/}}

\paragraph{Compatibility with Evolving Standards}
Long-term projects benefit from upgradable contracts to ensure compliance with industry standards or regulatory shifts. This flexibility allows adjustments that maintain compatibility without requiring users to interact with a new contract address, thereby enhancing usability and user retention.\footnote{\url{https://archive.trufflesuite.com/post/upgrading-smart-contracts-should-you-do-it-and-how}}

\subsubsection{Risks of Upgradeability}
\paragraph{Challenges to Immutability and User Trust}
The ability to modify contracts after deployment directly challenges the immutability that users expect in blockchain systems. Users often rely on the assurance that a contract will perform exactly as initially deployed. However, with upgradeability, users must trust the developers or administrators managing the upgrade functions. This added layer of trust can compromise the perceived neutrality and trustworthiness of the contract, as users may fear unauthorized or poorly managed changes.\footnote{\url{https://hacken.io/research/upgrading-smart-contracts-explanation-security-concerns}}

\paragraph{Security Vulnerabilities in Delegatecall and Storage Collisions}
Upgradeable contracts that rely on \texttt{delegatecall} (proxy-based approaches) inherit certain risks, including storage collisions and unauthorized access. A well-known example is the Audius hack, where a storage collision within a proxy contract allowed attackers to manipulate contract settings and access treasury funds. This incident underscores the need for precise storage layout management to avoid unauthorized actions, which can jeopardize contract security and disrupt user trust.\footnote{\url{https://audius.co/}}

\paragraph{Irreversible Bricking Through Partial Immutability}
Partial immutability, or "bricking," is a mechanism by which developers remove upgradeability from a contract, rendering its state permanent. This approach can build user trust by guaranteeing that a contract will become immutable after achieving stability. However, improper or premature use of bricking can lock out essential functions or block security updates, presenting a significant risk.\footnote{\url{https://ethereum.org/en/developers/docs/smart-contracts/upgradeability}}

\subsubsection{Best Practices for Mitigating Risks in Upgradeable Contracts}
To manage the security and trust challenges that accompany upgradeable contracts, projects adopt a series of best practices:
\begin{itemize}
    \item \textbf{Decentralized Governance for Security and Transparency:} Implementing decentralized governance models, such as multi-signature wallets and token-based voting, helps distribute control over upgrades. This minimizes the risks associated with a single point of failure, as decisions are collectively managed, ensuring no single party can unilaterally control upgrades.
    \item \textbf{Careful Use of Event Logging for Transparency:} Emitting on-chain events whenever an upgrade is proposed or completed improves visibility, allowing users to monitor contract changes in real time. These events enhance transparency and accountability by providing a clear audit trail for all upgrades, which reassures users and improves contract trustworthiness.\footnote{\url{https://metana.io/blog/upgrading-smart-contracts-heres-all-you-need-to-know}}
    \item \textbf{Use of Timelocks and Adaptive Control Mechanisms:} Timelocks provides a delay between when an upgrade is proposed and when it takes effect, giving users time to review and, if necessary, exit the contract. Adaptive timelocks, which adjust based on the severity or type of upgrade, can optimize the balance between security and responsiveness, especially in high-value contracts where time for community review is critical.\footnote{\url{https://compound.finance/governance}}
\end{itemize}

\subsubsection{Future Research Directions}
Despite the prevalent use of upgradeable smart contracts, limited empirical research addresses the impact of upgradeability on user trust, security, and long-term project sustainability. Given the growing importance of these mechanisms in DeFi and blockchain applications, future studies could yield critical insights into best practices and improvement areas:
\begin{itemize}
    \item \textbf{Empirical Studies on User Trust in Upgradeable Contracts:} To understand how users perceive and interact with upgradeable versus immutable contracts, empirical research should explore user attitudes toward trust and control in different upgradeability models. Surveys and behavioral studies assessing user willingness to engage with upgradeable contracts can inform developers on the transparency and control mechanisms users find most reassuring.
    \item \textbf{Impact of Upgradeability on Contract Security and Risk Mitigation:} As security incidents continue to highlight vulnerabilities in upgradeable systems, empirical studies focused on specific risks, such as \texttt{delegatecall} exploits and storage collisions, are needed. Case studies on security incidents like the Audius and Wormhole hacks can guide the development of risk assessment frameworks, ensuring that upgradeability mechanisms enhance security without compromising functionality.\footnote{\url{https://wormhole.com/}}
\end{itemize}

\section{Threats to Validity}
\label{sec:threats}
While our study aims for thoroughness and accuracy, potential internal threats to validity could impact the reliability of our findings. Addressing these threats is crucial to understanding the limitations of our research and the measures taken to mitigate potential biases. Below, we discuss the main types of validity threats: internal validity, external validity, construct validity, and reliability, highlighting specific concerns and the strategies used to manage them.

Internal validity in our study faces potential threats primarily due to the inclusion of both academic and grey literature. Grey literature, such as technical blogs and community forums, varies in credibility and quality, which could influence our conclusions. Although we employed systematic search strategies and predefined inclusion and exclusion criteria, the subjective nature of selecting grey literature may have introduced bias. We attempted to reduce this risk by conducting quality assessments using established frameworks and involving multiple reviewers in the selection process. Additionally, data extraction and thematic analysis are subject to interpretation, which may introduce bias. To mitigate this, we had multiple reviewers extract data independently and resolve discrepancies through consensus meetings, alongside calculating inter-rater reliability to ensure consistency. However, the inherent subjectivity of qualitative research may still affect the categorization and evaluation of upgrade approaches.

External validity is a concern as our focus on Ethereum smart contracts may limit our findings' generalizability to other blockchain platforms with different architectures, consensus mechanisms, or programming languages. This specialization could restrict the broader applicability of our conclusions. Moreover, while we conducted extensive searches across various databases and utilized citation tracking to minimize publication bias, the dynamic nature of the field may still influence the representation and emphasis of specific upgrade approaches, potentially affecting the balance of our analysis.

Construct validity in our study could be affected by the varied credibility of grey literature. Despite systematic methods and predefined criteria for inclusion, the subjective nature of selecting grey literature sources may impact how well the study captures the broader landscape of smart contract upgradeability. To address this, we performed quality assessments to maintain a consistent standard, though the variability in grey literature remains a consideration that could influence the quality and depth of our findings.

Reliability, or the consistency of our findings, is affected by the use of grey literature, which may not always be archived or easily accessible in the future. To improve reproducibility, we documented the output of these sources in detail and set clear criteria for selecting and assessing them. This documentation helps future researchers replicate or expand on our work. During our study, we encountered instances where some sources were no longer available online, which led to their exclusion from our analysis. Furthermore, the rapid pace of change in blockchain technology also affects the reliability of our study. New upgrade mechanisms or variations may have emerged after our knowledge cutoff in July 2024, which could influence the relevance of our findings. Regular updates and ongoing research are necessary to ensure our analysis remains up-to-date. Researchers and practitioners should consider the timing of our study when applying its conclusions to future projects.
\section{Conclusion} 
\label{sec:conclusion}
Smart contract upgradeability is critical to blockchain development, addressing the need for flexibility and adaptability in decentralized applications. Through our Multivocal Literature Review, we have systematically classified and analyzed the existing approaches to smart contract upgrades, bridging the gap between academic research and industry practices. Our study identifies two primary categories of upgrade mechanisms: full upgrade approaches, which involve redeploying contracts, and partial upgrade approaches, which modify specific components while preserving the contract state. We have provided a unified terminology to standardize the discussion of these methods.
We highlight the trade-offs inherent in selecting an upgrade mechanism by evaluating each approach against software quality attributes: complexity, flexibility, efficiency, security, and usability. For instance, proxy-based approaches offer moderate flexibility and maintain address consistency but introduce complexity and potential security risks. Data separation methods simplify storage management but may impact interaction simplicity.
Our analysis underscores the importance of considering project-specific requirements when selecting an upgrade approach. Developers must balance flexibility and adaptability with the principles of security and user trust inherent in blockchain technology. Furthermore, our discussion of governance models and lifecycle management emphasizes the role of transparent and robust governance in maintaining user trust during contract upgrades.
Future research should focus on empirical studies that evaluate the real-world impact of different upgrade mechanisms on security and user trust and the development of tools and standards that facilitate secure and efficient smart contract upgrades. By advancing the understanding of smart contract upgradeability, we aim to contribute to developing more resilient and adaptable blockchain applications.
\appendix
\section{Collected Data from Initial Grey Literature Search}
\label{app:A}
Details of all grey literature sources identified during the initial search phase, including those reviewed but not selected, are available online via \href{https://drive.google.com/file/d/15vZXjvfhoH7i2kyFt3TEOWvpprnATDni/view?usp=drive_link}{this link}.

\section{Collected Data from Initial Academic Literature Search}
\label{app:B}
Details of all academic literature sources identified during the initial search phase, including those reviewed but not included in the final analysis, are available online via \href{https://drive.google.com/file/d/1QZBqS3z74LmeOmasAwZSNSymVb0gjQow/view?usp=drive_link}{this link}.

\section{Detailed Inclusion and Exclusion Process}
\label{app:C}
The full list of academic and grey literature sources (both included and excluded), including reviewers' decisions, and reasons for inclusion or exclusion, is available online at \href{https://drive.google.com/file/d/1jOhsJypnTp8jiku7yrKN139MAe6D6Yz1/view?usp=drive_link}{this link}. 

\section{Comprehensive List of Included Sources}
\label{app:D}
List of all sources included in the study, both academic and grey literature, with details such as title, author, and year of publication are available online via \href{https://docs.google.com/spreadsheets/d/1oseEWkfTV-Idp4dzHmGGdMimSvs7A4lM/edit?usp=drive_link&ouid=108660595229926781828&rtpof=true&sd=true}{this link}. 

\section{Quality Assessment of Grey Literature Sources}
\label{app:E}
Details about the quality scores for all assessed grey literature sources, including the breakdown of scores for each criterion are available online via \href{https://docs.google.com/spreadsheets/d/1X5fMRInV2k-ubybSui0b991YUaemYwwf/edit?usp=drive_link&ouid=108660595229926781828&rtpof=true&sd=true}{this link}. 

\section{Pilot Study Results for Data Extraction Consistency}
\label{app:F}
Summary of the pilot study results are available online via \href{https://docs.google.com/spreadsheets/d/1IDyDpikTTxsYEo5Jm8EwSr0b_IAGc92Y/edit?usp=drive_link&ouid=108660595229926781828&rtpof=true&sd=true}{this link}.

\section{Standardized Data Extraction Form Template}
\label{app:G}
The template used for extracting data from sources, outlining columns and definitions for each data point is available online via \href{https://docs.google.com/spreadsheets/d/1vzoyTo99I_kOHyGLaTHh__NyvGJBVL38/edit?usp=drive_link&ouid=108660595229926781828&rtpof=true&sd=true}{this link}.

\section{Full Extracted Data from Included Sources}
\label{app:H}
Spreadsheet summarizing the extracted data from all included sources, covering their characteristics and identified benefits or limitations are available online via \href{https://docs.google.com/spreadsheets/d/1qVA8i5TJghQrNqfDsse8PnbywplFqqur/edit?usp=drive_link&ouid=108660595229926781828&rtpof=true&sd=true}{this link}. 

\section{Data Synthesis for RQ1}
\label{app:RQ1}
Overview of the thematic analysis conducted for RQ1, including initial coding and classification of smart contract upgrade approaches into Full and Partial Upgrade categories , is available online via \href{https://drive.google.com/file/d/1BcNnWleXANBDyG1ofVzx8no0WbMT1xoG/view?usp=drive_link}{this link}.

\section{Data Synthesis for RQ2}
\label{app:RQ2}
The framework analysis for RQ2 outlines data mapping to core components (address, logic, storage, execution flow), indexing, and thematic categorization. Further resources, including the full framework and analysis process, are accessible online via \href{https://drive.google.com/file/d/1oEQms2RnHwsfXcTEL21JI1uxXEfYWNBv/view?usp=drive_link}{this link}.

\section{Data Synthesis for RQ3}
\label{app:RQ3}
The thematic analysis steps for RQ3 include independent coding of benefits and limitations, card-sorting for theme grouping, and alignment with ISO/IEC 25010 quality criteria. Detailed steps of the analysis is available online via \href{https://drive.google.com/file/d/1kpKvdmIX-kT2GZkFifo2Po-QiWC338Fl/view?usp=drive_link}{this link}.

\section{Mapping of Standardized Terms to Literature}
\label{app:I}
A comprehensive table mapping various terms in the literature to the standardized terminology used in this study for consistency and cross-referencing is available online via \href{https://docs.google.com/spreadsheets/d/1NITpU3gpluTEDwFoYrTYMrf1qoegIVaA/edit?usp=drive_link&ouid=108660595229926781828&rtpof=true&sd=true}{this link}.





\bibliographystyle{ieeetr}
\bibliography{main.bib}
\end{document}